\documentclass[journal]{IEEEtran}
\usepackage{generic,hyperref,color}
\usepackage{cite}
\usepackage{amsmath,amssymb,amsfonts}
\usepackage{algorithmic}
\usepackage[ruled]{algorithm2e}
\usepackage{graphicx}
\usepackage{textcomp}
\usepackage{dsfont}
\usepackage{bbold}
\usepackage{tikz}
\usepackage{xcolor}
\usepackage{verbatim}
\usetikzlibrary{arrows}

\newtheorem{lemma}{Lemma}
\newtheorem{proposition}{Proposition}
\newtheorem{definition}{Definition}
\newtheorem{theorem}{Theorem}
\newtheorem{corollary}{Corollary}
\newtheorem{example}{Example}

\newtheorem{remark}{Remark}

\DeclareMathOperator{\dist}{dist}
\DeclareMathOperator*{\argmax}{arg\,max}

\newcommand{\mc}{\mathcal}

\newcommand{\ds}{\displaystyle}

\newcommand{\R}{\mathds{R}}

\newcommand{\be}{\begin{equation}}
\newcommand{\ee}{\end{equation}}
\newcommand{\ba}{\begin{array}}
\newcommand{\ea}{\end{array}}

\newcommand{\wz}{x^*}
\newcommand{\eps}{\varepsilon}
\newcommand{\tx}{\tilde{x}}

\newcommand{\tcb}{\textcolor{black}}
\newcommand{\1}{\mathbb{1}}
\newcommand{\ones}{\mathbf{1}}

\newcommand{\de}{\mathrm{d}}
\newcommand{\sign}{\mathrm{sgn}}
\newcommand{\se}{\text{ if }}

\def\qed{\hfill \vrule height 7pt width 7pt depth 0pt\medskip}

\begin{document}
\title{\tcb{On the Stability of the Logit Dynamics in Population Games}}
\author{Leonardo Cianfanelli and Giacomo Como, 
\IEEEmembership{Member, IEEE}
\thanks{Some of the results appeared in a preliminary form in \cite{cianfanelli2019stability,cianfanelli2022stability}.}
\thanks{This research received partial support from the MIUR  Research Project PRIN 2017 ``Advanced Network Control of Future Smart Grids'' (http://vectors.dieti.unina.it), and by the {\it Compagnia di San Paolo} through a Joint Research Project and project ``SMAILE-- Simple Methods for Artificial Intelligence Learning and Education''.}
\thanks{Leonardo Cianfanelli and Giacomo Como are with 
Dipartimento di Scienze Matematiche, Politecnico di Torino. Giacomo Como is also with  the Department of Automatic Control, Lund University (e-mail: leonardo.cianfanelli@polito.it, giacomo.como@polito.it, webpage: https://sites.google.com/view/leonardo-cianfanelli/, https://staff.polito.it/giacomo.como/)}
}

\maketitle

\begin{abstract}
We study the asymptotic stability of the logit evolutionary dynamics in population games, \tcb{possibly with multiple heterogenous populations}. For
general population games, we prove that, on the one hand, strict Nash
equilibria are asymptotically stable under the logit dynamics
for low enough noise levels, on the other hand, a globally exponentially
stable logit equilibrium exists for sufficiently large noise levels. This suggests the
emergence of bifurcations in population games admitting multiple strict Nash equilibria, as observed in numerous examples.
We then provide sufficient conditions on the population game structure for the existence of globally
asymptotically stable logit equilibria for every noise level. The considered class of monotone separable games finds applications, e.g., in routing games on series compositions of networks with parallel routes when there are multiple populations of users that differ in the reward functions.
\end{abstract}


\section{Introduction}
Population games \cite{blume2018population,sandholm2010population} provide a powerful mathematical framework to model strategic \tcb{interactions} in large-scale multi-agent systems.  
In this class of games, the set of players is divided into a finite number of populations, which may differ in size, in the available action sets, and in the reward functions modeling the agents' preferences. Moreover, the interactions among players are anonymous, namely, the reward of a player depends only on her action and on the aggregate action distribution across the populations. Population games find applications in many socio-technical systems, particularly, in the study of congestion in traffic networks, of the emergence of conventions, norms, and institutions, as well as of public goods and externalities. They have also been extensively studied in the context of evolutionary biology \cite{smith1974theory}. 

Population games can be studied with either atomic (finite) player set \cite{blume2018population} or non-atomic (continuous) player set \cite{sandholm2010population}: in this paper, we shall focus on the latter case.
The equilibrium analysis of continuous population games has attracted a large amount of attention in the past decades.   
The existence, uniqueness, and efficiency of Nash equilibria in population games have been widely studied, in particular in the special case of routing games \cite{schmeidler1973equilibrium,milchtaich2005topological,roughgarden2002bad,roughgarden2003price,roughgarden2004bounding,patriksson2015traffic,dafermos1973toll}. However, the notion of equilibrium is completely static, and does not take into account the dynamics of the users' decisions. On the other hand, \emph{evolutionary game theory} \cite{weibull1997evolutionary,hofbauer2003evolutionary,sandholm2010population} studies the collective behaviors emerging in population games when the players modify their actions over time according to certain revision protocols. The resulting evolutions of the distribution of actions across the population are described as dynamical systems, to be referred to as \emph{evolutionary dynamics}. 
Issues pertaining convergence to and stability of equilibria, or equilibrium selection, are crucial in the study of evolutionary dynamics in population games, both from a theoretical perspective and for the applications.

In this paper, we focus on the logit dynamics, a special case of evolutionary dynamics modeling the behavior of agents that aim at selecting optimal actions, but due to some noise may sometimes adopt suboptimal ones \cite{blume1993statistical,marden2012revisiting,sandholm2010population,mckelvey1995quantal}. The noise may represent errors in estimating the rewards or in implementing the best response actions, resulting in smooth dynamics that are often more easily tractable from a mathematical viewpoint than the best response dynamics.

Our contribution is twofold.
First, we analyze the asymptotic behavior of the logit dynamics for low noise levels in arbitrary population games, proving that (Theorem \ref{thm:logit}) for every strict Nash equilibrium (i.e., an equilibrium where all players of a population play the same action, which gives a strictly higher reward to players in that population than any of their available alternatives) there is a curve of asymptotically stable logit equilibria that approach the strict Nash equilibrium as the noise level vanishes.
This result is complemented by Theorem \ref{thm:noise}, establishing the existence of a globally exponentially stable logit equilibrium for large enough noise levels. Both these results hold true for every population game. For population games admitting multiple strict Nash equilibria, they suggest the emergence of bifurcations for the logit dynamics with respect to the noise level. Such bifurcations are indeed observed in numerous examples.

Second, we determine novel sufficient conditions for global asymptotic stability of Nash equilibria in population games under the logit dynamics. Specifically, we prove that (Theorem~\ref{thm:conv_global}), for a class of population games to be referred to as \emph{monotone separable}, there exists a globally asymptotically stable logit equilibrium for every noise level, and that such logit equilibria form a curve approaching a Nash equilibrium of the population game in the vanishing noise limit. Notably, we provide explicit examples of monotone separable population games for which global asymptotic stability of logit equilibria does not follow from any of the currently available results in the literature.

We then show how our results can be applied to heterogeneous routing games, where the populations differ one from another in their reward functions \cite{dafermos1972traffic, farokhi2013heterogeneous}. Indeed, while in homogeneous routing games Nash equilibria are known to be globally asymptotically stable under a large class of evolutionary dynamics due to the existence of a \tcb{concave} potential function \cite{sandholm2010population}, the stability analysis of evolutionary dynamics ---including the logit dynamics--- in heterogeneous routing games is a largely open issue. In particular, we show that Theorem \ref{thm:conv_global} implies the global asymptotic stability, under the logit dynamics, of Nash equilibria of heterogeneous routing games when the routing network is a series composition of parallel graphs. 
In fact, our global stability results are proved to hold true for a class of heterogeneous routing games where the cost function associated to each link of the network is a function of a weighted aggregate flow over the link, thus extending the results in \cite{cianfanelli2019stability}. This generalization finds applications, e.g., in mixed autonomy routing games \cite{mehr2019will,arcak2020dissipativity}.
To the best of our knowledge, this is the first global stability result for the logit dynamics in heterogeneous routing games that do not admit a potential function, besides our preliminary results published in \cite{cianfanelli2019stability}. 


Our results on the asymptotic stability of logit equilibria in population games complement and extend some of those already available in the literature.
In particular, on the one hand,  \cite{hofbauer2001nash} and \cite[Theorem 8.4.6]{sandholm2010population} establish the asymptotic stability of interior evolutionary stable states (ESS) under the logit dynamics. ESSs constitute a  (possibly empty) subset of the set of Nash equilibria of a population game that encompass strict Nash equilibria. However, strict Nash equilibria are never interior, hence the asymptotic stability of strict Nash equilibria characterized in our Theorem \ref{thm:logit} is not implied by the asymptotic stability of interior ESSs, nor vice versa. 

On the other hand, global asymptotic stability of logit equilibria in population games is established in the literature for potential, supermodular, and (weighted) contractive\footnote{Contractive games were originally termed \emph{stable} games, see e.g., \cite{sandholm2010population}.} population games \cite{hofbauer2009stable,hofbauer2007evolution,sandholm2010population,arcak2020dissipativity}. Moreover, global asymptotic stability of perturbed equilibria has been proven for games with interior evolutionary stable states and zero-sum games under stochastic fictitious play \cite{hofbauer2002global}, a class of learning dynamics that combine perturbed best response choices (such as the logit choice) with fictitious play. We remark that none of these classes of games encompass monotone separable games.

It is also worth mentioning a recent branch of the control systems literature analyzing the stability of evolutionary dynamics by leveraging tools from passivity theory \cite{fox2013population,gao2020passivity,martinez2023distributed}. Indeed, many evolutionary dynamics are known to be $\delta$-passive \cite{park2019population}, including the logit dynamics, which are known to be $\delta$-passive with surplus equal to the noise level \cite{park2018passivity}. Combining various notions of passivity of evolutionary dynamics with the anti-passivity of many population games it is possible to derive global stability results. These techniques find applications, e.g., in contractive games, which are known to be $\delta$-antipassive, and in weighted-contractive games \cite{arcak2020dissipativity}. This theory allows to prove global stability results also in more complex scenarios where the rewards have themselves some dynamics instead of being a static function of the configuration of the game, and agents of different populations revise their actions according to different evolutionary dynamics \cite{park2019population,park2019payoff}. 

In this paper we adopt another approach based on the theory of contractive systems \cite{FB-CTDS}. However, these notions of passivity are closely related to our work. For instance, the global asymptotic stability of the logit dynamics in the large noise limit (Theorem \ref{thm:noise}) may be alternatively proved by using the passivity properties of the logit dynamic.

The rest of this paper is organized as follows. In Section \ref{sec:model}, we introduce population games, the logit dynamics, and provide some motivating examples. In Section \ref{sec:results}, we provide our main results on the stability of the logit dynamics in population games. Section \ref{sec:routing} discusses the behavior of the logit dynamics in routing games. 
Finally, Section \ref{sec:evo_conclusion} summarizes our work and presents some possible future research lines.

\subsection*{Notation}
Let $\mathds{R}$ and $\mathds{R}_{+}$ denote the set of real numbers and non-negative real numbers, respectively. For a finite set $\mc{A}$, its cardinality is denoted by $|\mc{A}|$, while $\mathds{R}^{\mc{A}}$ denotes the space of real-valued \tcb{column} vectors, whose entries are indexed by the elements of $\mc{A}$. Let $\ones$ denote the all-$1$ vector, while $\delta^{(i)}$ denotes the vector with $1$ in position $i$ and $0$ in all other positions, whose size may be deduced from the context.
The indicator function of a set $\mc A$ is denoted by  $\1_{\mc A}$, so that $\1_{\mc A}(a)=1$ if $a$ belongs to $\mc A$  and $\1_{\mc A}(a)=0$ if $a$ does not belong to $\mc A$.
\tcb{
The $l_1$-norm of a vector $x$ in $\R^{\mc A}$ is denoted by $$||x|| = \sum_{a\in\mc A} |x_{a}|\,,$$ and the corresponding distance between $x$ and a set $\mc Y \subseteq \R^{\mc A}$} is defined as
$$\dist(x,\mc Y)=\inf\{||x-y||:\,y\in\mc Y\}\,.$$
For $\eps>0$, 
$$\mc B_{\eps}(\mc Y)=\{x\in \tcb{\R^{\mc A}}:\,\dist(x,\mc Y)<\eps\}$$
denotes the set of points at distance less than $\eps$ from $\mc Y$.

\section{Model}
\label{sec:model}
In this section, we first introduce population games and the logit dynamics. We then recall a few known results on the asymptotic stability of logit equilibria in certain classes of population games. Finally, we present some motivating examples illustrating asymptotic behaviors that will be the object of our study in the following sections.

\subsection{Population games}
Population games model anonymous strategic interactions among very large populations of agents \cite{sandholm2010population}. In this class of games, the agent set is assumed to form a continuum and is divided into a non-empty finite set of populations that may differ one from another in size, available action set, and reward functions. 
Agents within the same population are indistinguishable and can choose among a finite set of actions, aiming at maximizing a reward that is a continuous function of the aggregate action distribution.  
Formally, a population game is a quadruple \be\label{def:pop-game}(\mc{P},\{\mc{S}_p\}_{p \in \mc P},v,r)\,,\ee where:
	\begin{itemize}
		\item $\mc{P}$ is a finite set of populations, together forming a society;
		\item $v$ in \tcb{$(0,+\infty)^\mc{P}$} is a \tcb{positive vector, whose $p$'th entry $v_p$ represents the positive size of population $p$, for $p$ in $\mc P$};
		\item $\mc S_p$ is the set of actions available to players in population $p$ in $\mc P$, and 
		$$\mc S=\bigcup_{p\in\mc P}\mc S_p\,,$$ denotes the set of all available actions in the society; 		 
		\item \tcb{a (society) configuration is a tuple $x$ in $\R_+^{\mc S \times \mc P}$ such that
		\be\label{eq:X}
		\ba{rcll}
		\sum_{i\in\mc S}x_{ip} & = & v_p\,, \ \ & \forall p \in \mc P\,,\\[6pt]
		x_{ip} & = & 0\,, \ &\forall i \notin \mc S_p\,,  \forall p \in \mc P\,.
		\ea\ee
		\item the $(i,p)$-th entry $x_{ip}$ of a configuration $x$ represents the mass of players of population $p$ playing action $i$; 
		\item the space of configurations is denoted by}
		$$
		\tcb{\mc X =\left\{x \in \R_+^{\mc S \times \mc P}: \eqref{eq:X}\right\}};
		$$
		\item the reward function $$r: \mc{X} \to \tcb{\mathds{R}^{\mc{S} \times \mc P}}$$ is Lipschitz \tcb{on $\mc X$ and continuously differentiable on the interior of $\mc X$}, and such that $r_{ip}(x)$ represents the reward that every player of population $p$ in $\mc P$ gets choosing action $i$ in $\mc S$ in configuration  $x$ in $\mc X$. 
	\end{itemize}

	A \emph{Nash equilibrium} for a population game \eqref{def:pop-game} is a configuration $x$ in $\mc{X}$ such that, for every population $p$ in $\mc P$ and available action $i$ in $\mc{S}_p$,
	\begin{equation}\label{eq:nash}
		x^*_{ip}>0\qquad \implies\qquad r_{ip}(x^*)\ge r_{jp}(x^*)\,, \quad \forall j \in \mc{S}_p.
	\end{equation}
In plain words, a Nash equilibrium is a configuration such that every action played by a positive fraction of players in a population yields maximal reward among all the actions available to that population. In fact, an equivalent condition for a configuration $x$ to be a Nash equilibrium is that 
\be\label{eq:ess}
\sum_{i \in \mc S_p} (x_{ip}-y_{ip})r_{ip}(x) \ge 0\,,\qquad \forall p \in \mc P\,,
\ee
for every other configuration $y$ in $\mc X$. 
We shall denote by $$\mc{X}^*=\{x^*\in\mc X:\,\eqref{eq:nash}\}\,,$$ the set of Nash equilibria of a population game. 

We shall refer to a Nash equilibrium $x^*$ in $\mc X^*$ as:
\begin{itemize}
\item \emph{strict} if, for every $p$ in $\mc P$,
$$
	x_{ip}^*>0 \implies r_{ip}(x^*) > r_{jp}(x^*)\,, \quad \forall j \in \mc{S}_p \setminus \{i\}\,;
$$
\item \emph{monomorphic} if a single action is played in each population, i.e., for every $p$ in $\mc P$ there exists $s_p$ in $\mc S_p$ such that
\be\label{eq:monomorphic}x^*_{ip} =\left\{\ba{rcl}v_p&\se& i=s_p\,,\\[2pt]0&\se&i\ne s_p\,;\ea\right. \ee
\item \emph{isolated} if 
there exists some $\epsilon>0$ such that $$\mc B_\epsilon (x^*) \cap \mc{X}^* =\{x^*\}\,.$$
\end{itemize}

We let $\mc X^*_s\subseteq\mc X^*$ denote the set of strict Nash equilibria.  Then, the following result holds true. 
\begin{proposition}\label{prop:existenceNash}
	For every population game \eqref{def:pop-game}: 
	\begin{enumerate}
		\item[(i)] the set of Nash equilibria $\mc X^*$ is nonempty and compact; 
		\item[(ii)] every strict Nash equilibrium is monomorphic; 
		\item[(iii)] every strict Nash equilibrium is isolated. 
	\end{enumerate}
\end{proposition}
\begin{IEEEproof} 
	 See Appendix \ref{app:prop1}.\end{IEEEproof}\medskip
Notice that, while Proposition \ref{prop:existenceNash}(i) ensures the existence of Nash equilibria for every population game, the set $\mc X^*_s$ of strict Nash equilibria of a population game may be empty. 

\begin{example}[Binary coordination]	Consider a single-population game with unitary mass of players $v=1$, binary action set $\mc S = \{1,2\}$, and reward function
		\be\label{cost-12coordination}r_1(x) = x_1\,,\qquad r_2(x)=\gamma x_2\,,\ee
		where $\gamma>0$ is a parameter. 
		For every $\gamma>0$, the set of Nash equilibria is $\mc X^*=\{\delta^{(1)},\delta^{(2)},x^{*}\}$, 
		where
$$ \delta^{(1)} = (1,0)\,,\qquad \delta^{(2)} = (0,1)\,,\qquad x^{*} = \left(\frac{\gamma}{1+\gamma},\frac{1}{1+\gamma}\right)\,.$$
		Moreover, the Nash equilibria $\delta^{(1)}$ and $\delta^{(2)}$ are strict, while the Nash equilibrium $x^{*}$ is not, i.e., $\mc X^*_s=\{\delta^{(1)},\delta^{(2)}\}$.
\end{example}
\begin{example}[Generalized Rock-Scissor-Paper]
Consider a single-population game with unitary mass of players $v=1$, ternary action set $\mc S=\{1,2,3\}$, and reward function
\be\label{cost-RSP}r(x) =Rx\,,\qquad R=\left(\ba{ccc}\gamma&-1&1\\ 1&\gamma&-1\\-1&1&\gamma\ea\right)\,,\ee
where $\gamma\ge0$ is a parameter. 
Then, $x^*=(1/3,1/3,1/3)$ is a non-strict Nash equilibrium for every $\gamma\ge0$. 
Moreover, for $\gamma<1$ there are no other Nash equilibria. 
On the other hand, for $\gamma\ge1$ there are three monomorphic Nash equilibria  
$$ \delta^{(1)} = (1,0,0)\,,\qquad \delta^{(2)} = (0,1,0)\,,\qquad \delta^{(3)} = \left(0,0,1\right)\,,$$
and they are all strict if and only if $\gamma>1$. 
Therefore, the set of Nash equilibria is 
$$\mc X^*=\left\{\ba{lcl}\{x^*\}&\se&0\le\gamma<1\,,\\[3pt]\left\{x^*,\delta^{(1)},\delta^{(2)},\delta^{(3)} \right\}&\se&\gamma \ge 1\,,\ea\right.$$
whereas the set of strict Nash equilibria is $$\mc X^*_s=\left\{\ba{lcl}\emptyset&\se&0\le\gamma\le1\,,\\[3pt]\left\{\delta^{(1)},\delta^{(2)},\delta^{(3)} \right\}&\se&\gamma>1\,.\ea\right.$$

\end{example}

\begin{example}
Consider a population game with two populations, both of unitary mass, i.e.,  $\mc P=\{1,2\}$ and  $v = \ones$. Let the action sets be $\mc S_1=\mc S_2=\mc S=\{1,2,3\}$ and the reward function be defined by 
\be\label{eq:reward_cianfanelli}
\!
\ba{ll}r_{11}(x)=-(x_{11}+x_{12})\,,& r_{12}(x)=-2\,,\\[3pt]
r_{21}(x) = -1 \,,& r_{22}(x) = -(x_{21}+x_{22})\,,\\[3pt]
r_{31}(x) = -2 \,,& r_{32}(x) = -1\,. \ea
\ee
		This population game admits a unique Nash equilibrium \tcb{$x^*$, which has entries 
		$$x^*_{11}=x^*_{22}=1\,,\qquad x^*_{21}=x^*_{12}=x^*_{31}=x^*_{32}=0\,.$$}
		Such Nash equilibrium is monomorphic but not strict. Hence, $\mc X^*=\{x^*\}$ and $\mc X^*_s=\emptyset$.
\end{example}
\subsection{Logit dynamics in population games}
\label{sec:logit}
\tcb{Given a population game \eqref{def:pop-game} and a noise level $\eta>0$, we define the logit choice function $F:\mc X \times (0,+\infty) \to \R^{\mc S\times\mc P}$ by
\be\label{def:G}			
F_{ip}(x,\eta) =
\1_{\mc S_p}(i)\frac{v_p\exp(r_{ip}(x)/\eta)}{\sum_{j \in \mathcal{S}_p} \exp(r_{jp}(x)/\eta)}\,,
\ee
for every configuration $x$ in $\mc X$, population $p$ in $\mc P$, and action  $i$ in $\mc S$. }
\tcb{The logit choice function is also known in the literature as soft-max, perturbed best response, smoothed best response, or quantal response  \cite{mckelvey1995quantal} function and is widely used in discrete choice models \cite{Brock.Durlauf:2001}. 
In particular, the right-hand side of equation \eqref{def:G} can be read as the probability of choosing action $i$ in $\mc S$ for a player in population $p$  whose objective is to maximize the expected value of the sum of the reward plus a stochastic perturbation with Gumbel probability distribution (see, e.g., \cite[Section 6.2.3]{sandholm2010population}). }

The following statement gathers some fundamental properties of the logit choice function that will prove useful in the rest of the paper.
\tcb{\begin{lemma}\label{lemma0}
Consider a population game \eqref{def:pop-game}. Then:
\begin{enumerate}
\item[(i)] for every noise level $\eta>0$, 
\be\label{eq:Fx-x}
F(x,\eta) \in \mc X\,, \quad \forall x \in \mc X\,;
\ee
\item[(ii)] in the large noise limit, the logit choice function converges to a uniform distribution on the action set of each population, i.e., 
$$
\lim_{\eta \to +\infty}F_{ip}(x,\eta) = \1_{\mc S_p}(i) \frac{v_p}{|\mc S_p|}\,,\quad \forall i \in \mc S, p \in \mc P\,;
$$
\item[(iii)] in the small noise limit, the logit choice function converges to a uniform distribution on the best response set of each population, i.e., 
$$\lim_{\eta \to 0^+} F_{ip}(x,\eta) = \1_{\mc{S}^*_p(x)}(i)\frac{v_p}{|\mc S^*_p(x)|} \,, \quad \forall i \in \mc S, p \in \mc P\,,
$$ 
where $$
\mc{S}_p^*(x) = \argmax_{i \in \mc S_p} \ r_{ip}(x)
$$
is the set of optimal actions for players of population $p$ in configuration $x$;
\item[(iv)] for every noise level $\eta>0$, the function $x \mapsto F(x,\eta)$ is Lipschitz on $\mc X$ and continuously differentiable on the interior of $\mc X$;
\item[(v)] for every configuration $x$ in $\mc X$, the function $\eta \mapsto F(x,\eta)$ is continuously differentiable on $(0,+\infty)$;
\item[(vi)] for every noise level $\eta>0$, $x \mapsto F(x,\eta)$ admits at least one fixed point in $\mc X$.
\end{enumerate}
\end{lemma}}
\begin{IEEEproof}
	See Appendix \ref{app:0}.
\end{IEEEproof}\medskip

For a noise level $\eta>0$, the logit dynamic is the continuous-time dynamical system
\be\label{eq:logit}
\dot{x} = F(x,\eta)-x \,.\ee
The logit dynamics are a special case of evolutionary dynamics \cite{sandholm2010population}, i.e., continuous-time dynamical systems describing the evolution of plays in a population game as the players revise their decisions over time. The logit dynamics model agents that aim at selecting optimal actions, but due to some noise may sometimes make mistakes and take suboptimal actions. It can be interpreted as the mean-field limit (in the spirit of Kurtz's theorem \cite{Kurtz1981}) of the asynchronous noisy best response dynamic for finite strategic games \cite{blume1993statistical}. 

For a population game \eqref{def:pop-game} and a noise level $\eta>0$, let 
$$\mc X^{(\eta)}=\left\{x\in\mc X:\,F(x,\eta)=x\right\}\,,$$ 
denote the set of \emph{logit equilibria}, i.e., equilibrium points of the logit dynamic \eqref{eq:logit}. Let also 
$$\mc X^{(0)}\!=\!\left\{x\in\mc X: \exists \eta_n\!>\!0, \eta_n\!\!\stackrel{n\to+\infty}{\longrightarrow}\!0, x^{n}\!\in\!\mc X^{(\eta_n)}, x^{n}\!\stackrel{n\to+\infty}{\longrightarrow}\! x\right\}$$ denote the set of accumulation points of sequences of logit equilibria as the noise level vanishes, to be referred to as \emph{limit logit equilibria}. The following result is standard \cite{mckelvey1995quantal}.  
\begin{lemma}\label{lemma:existence-uniqueness}
Consider a population game \eqref{def:pop-game}. Then:
\begin{enumerate} 
\item[(i)] the ODE \eqref{eq:logit} defines a well-posed dynamical system on the configuration space $\mc X$ for every noise level $\eta>0$;
\item[(ii)] the set of logit equilibria $\mc X^{(\eta)}$ is nonempty and compact for every noise level $\eta>0$;
\item[(iii)] the set of limit logit equilibria $\mc X^{(0)}\subseteq\mc X^*$ is a nonempty compact subset of the set of Nash equilibria. 
\end{enumerate}
\end{lemma}
\begin{IEEEproof} See Appendix \ref{sec:proof-lemma:existence-uniqueness}.
\end{IEEEproof}	

\subsection{Asymptotic stability of logit equilibria}\label{sec:asymp-stability}

For a population game \eqref{def:pop-game} and a noise level $\eta>0$, a logit equilibrium $x^{(\eta)}$ in $\mc X^{(\eta)}$ is referred to as: 
\begin{itemize}
\item \emph{asymptotically stable} if, for every $\eps>0$, there exists $\delta>0$ such that, for every initial configuration $x(0)$ in $\mc B_{\delta}(x^{(\eta)})\cap\mc X$, the solution of the logit dynamic \eqref{eq:logit} with noise level $\eta>0$ is such that $x(t)$ belongs to $\mc B_{\eps}(x^{(\eta)})$ for every $t\ge0$ and 
\be\label{limit}
\lim_{t \to +\infty} x(t) = x^{(\eta)}\,;\ee
\item  \emph{globally asymptotically stable} if it is asymptotically stable and \eqref{limit} holds true for every initial condition $x(0)$ in $\mc X$; 
\item  \emph{globally exponentially stable} if it is asymptotically stable and there exists a constant $\alpha>0$ such that
$$
\tcb{||x(t)-x^{(\eta)}|| \le ||x(0)-x^{(\eta)}||} e^{-\alpha t}\,,\qquad \forall t \ge 0\,,
$$
for every initial configuration $x(0)$ in $\mc X$.
\end{itemize}

In the rest of the paper, we shall investigate the stability of the logit dynamics in population games. In particular, we shall determine sufficient conditions for the (global, exponential) asymptotic stability of logit equilibria in terms of both the structure of the population game and the noise level. 
We shall now recall some of the best known results on the asymptotic stability of logit equilibria that are available in the literature on population games. 

On the one hand, asymptotic stability of the logit dynamics for low noise levels is known to hold true for so-called \emph{evolutionary stable states (ESSs)}. Specifically, a configuration $x$ in $\mc X$ of a population game \eqref{def:pop-game} is a (Taylor) ESS \cite[Section 8.3.3]{sandholm2010population} if $x$ is a Nash equilibrium, and there exists $\eps>0$ such that
$$\sum_{p \in \mc P}\sum_{i \in \mc S_p}(y_{ip}-x_{ip})r_{ip}(y) < 0\,,$$ 
for every $y\ne x$ in $\mc B_\eps(x) \cap \mc X$ such that \eqref{eq:ess} holds true as an equality. 
For an ESS $x$ that lies in the interior of the configuration space $\mc X$, \cite[Theorem 8.4.6]{sandholm2010population} ensures that for some sufficiently small noise level $\eta>0$ and $\eps>0$, there exists a unique logit equilibrium $x^{(\eta)}$ in $\mc B_\eps(x)$, and that such $x^{(\eta)}$ is asymptotically stable. Moreover, $$\lim_{\eta \to 0^+} x^{(\eta)} = x\,,$$ hence \tcb{interior} ESSs are asymptotically stable limit logit equilibria. \tcb{Such a result may be proved by showing that
\be\label{eq:lya}
\ba{rcl}
\!\!\!	V_\eta(x) &\!\! = &\!\! \ds \eta \sum_{p \in \mc P} v_p \log\Big( \sum_{i \in \mc S_p} \exp(r_{ip}(x)/\eta)\Big) + \\[6pt]
& \!\! + &\!\! \ds \eta \sum_{p \in \mc P} \sum_{i \in \mc S_p} x_{ip} \log\frac{x_{ip}}{v_p}\ - \sum_{p \in \mc P}\sum_{i \in \mc S_p} x_{ip} r_{ip}(x)\,,
\ea
	\ee
is a local strict Lyapunov function for the logit dynamics in a neighborhood of an interior ESS.} 
It is worth pointing out that every strict Nash equilibrium $x$ in $\mc X_s^*$ is an ESS, but it is never interior by Proposition \ref{prop:existenceNash}(ii).

On the other hand, global asymptotic stability of logit equilibria is established in the literature for certain classes of population games having particular structure.
Particularly relevant are the cases of potential and (weighted) contractive games, as illustrated below. 
	A population game \eqref{def:pop-game} 
	is called \emph{potential} if there exists a differentiable function $\Phi : \mc{X} \to \mathds{R}$ such that
	\be\label{eq:potential}
			r_{ip}(x) - r_{jp}(x) = \frac{\partial \Phi(x)}{\partial x_{ip}} - \frac{\partial \Phi(x)}{\partial x_{jp}}\,,
	\ee
	for every $x$ in $\mc X$, $p$ in $\mc P$, and $i, j$ in $\mc S_p$. 
	When the reward function $r$ is differentiable, this is equivalent to (c.f.~\cite[Theorem 3.2.2, Exercise 3.2.4]{sandholm2010population})
	\be\label{eq:pot_games}
	\left(\!\frac{\partial}{\partial x_{hq}}\!-\!\frac{\partial}{\partial x_{kq}}\!\right)\!\left(r_{ip}\! -\! r_{jp}\right) = 		
		\!\left(
	\!\frac{\partial}{\partial x_{ip}}\!-\!\frac{\partial}{\partial x_{jp}}\!\right)\!\left(r_{hq}\! -\! r_{kq}\right),
	\ee
	for every $p$ and $q$ in $\mc P$, $i$ and $j$ in $\mc S_p$, and $h$ and $k$ in $\mc S_q$. 
	It then follows from \cite[Theorem 7.1.4]{sandholm2010population} that, for every noise level $\eta>0$, the logit dynamics in potential population games admit the strict Lyapunov function $$\tcb{V_{\eta}(x) = - \Phi(x) + \eta \sum_{p \in \mc P}\sum_{i \in \mc S_p} x_{ip} \log\frac{x_{ip}}{v_p}\,,}$$ whose set of stationary points in $\mc X$ coincides with the set of logit equilibria $\mc X^{(\eta)}$, so that $x(t)$ 
	converges to $\mc X^{(\eta)}$ as $t$ grows large. In particular, if the potential function of the population game $\Phi(x)$ is concave, then the Lyapunov function $V_\eta(x)$ is strictly convex for every noise level $\eta>0$, and $\mc X^{(\eta)}$ is a singleton that approaches the set of limit equilibria $\mc X^{(0)}$ in $\mc X^*$ as the noise level $\eta$ vanishes. 

A population game \eqref{def:pop-game} is called \emph{weighted-contractive} if there exists a vector of weights $\alpha$ in $\tcb{(0,+\infty)^\mc P}$ such that
	\be\label{contractive}
	\sum_{p\in\mc P}\sum_{i\in\mc S} \alpha_p (y_{ip}-x_{ip})(r_{ip}(y)-r_{ip}(x)) \le 0\,, \ee
	for every two configurations $x$ and $y$ in $\mathcal X$. If $\alpha = \ones$, weighted-contractive games are simply referred to as \emph{contractive}. It follows from \cite[Theorem 7.2.8]{sandholm2010population} and \cite{hofbauer2007evolution} that  the set of Nash equilibria $\mc X^*$  of a contractive population game is convex and that for every noise level $\eta>0$ there exists a globally asymptotically stable logit equilibrium  $x^{(\eta)}$ that approaches $\mc X^*$ in the vanishing noise limit. \tcb{The asymptotic stability of logit equilibria for contractive games may be proved by showing that \eqref{eq:lya} is a global  strict Lyapunov function for the logit dynamics, or, alternatively, it may be} proved using results from passivity theory for dynamical systems, see, e.g., \cite{park2019population}. Moreover, in weighted-contractive population games, the set $\mc{X}^{(\eta)}$ is known to be globally asymptotically stable under the logit dynamics for every noise level $\eta>0$ based on the $\delta$-passivity of the logit dynamics \cite[Corollary 1]{arcak2020dissipativity}.


\subsection{Examples} \label{sec:examples}
We conclude this section by looking at the behavior of the logit dynamics in the population games introduced in Examples \ref{ex:coordination}--\ref{ex:2pop2link}.
In particular, such examples will help us illustrating some limitations of the results recalled in Section \ref{sec:asymp-stability} in predicting the behavior of the logit dynamics in these games, thus motivating the novel results derived in Section \ref{sec:results}.

\begin{figure}
	\centering
	\includegraphics[width=6cm]{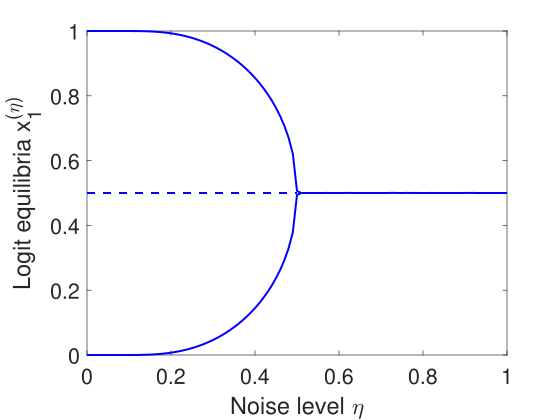}\\[2pt]
	\includegraphics[width=6cm]{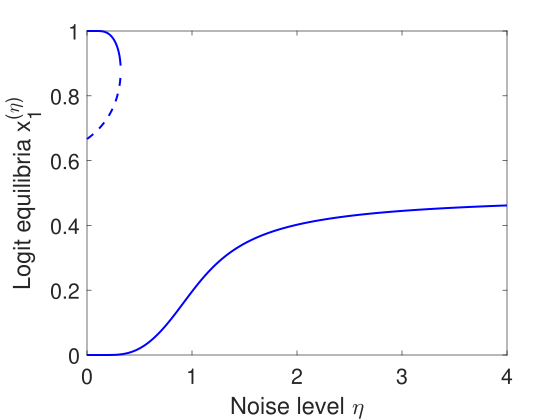}
	\caption{\label{fig:coordination} 
		Bifurcation diagrams of the logit dynamics in the population games of Example \ref{ex:coordination} in the special cases $\gamma=1$ (\emph{top}) and $\gamma=2$ (\emph{bottom}).}
\end{figure}

\begin{figure}
	\centering
	\includegraphics[width=4.35cm]{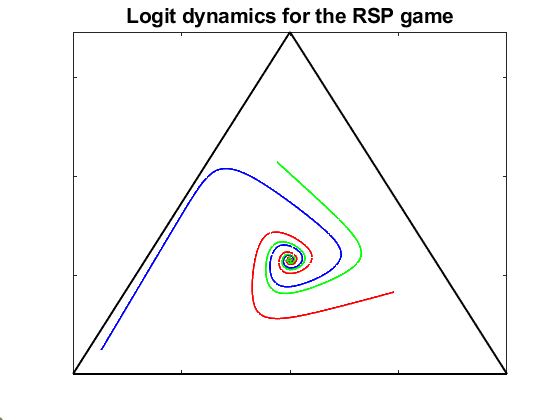}
	\includegraphics[width=4.35cm]{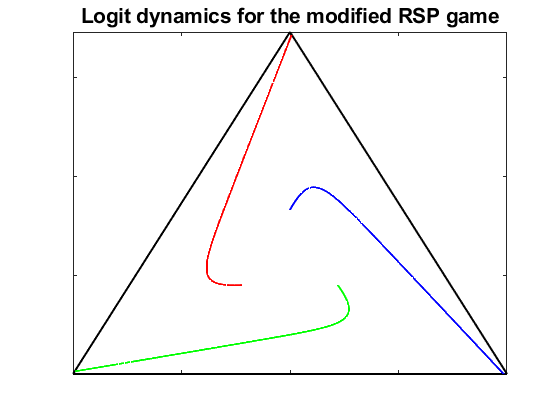}
	\includegraphics[width=4.35cm]{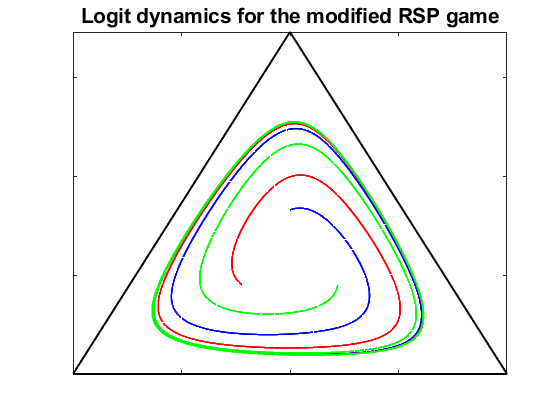}
	\includegraphics[width=4.35cm]{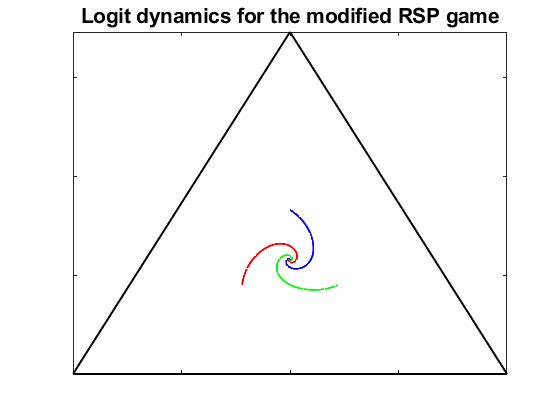}
	\caption{\label{fig:RSP} Solutions of the logit dynamics in the games of Example \ref{ex:RSP} for different values of $\gamma$ and noise levels $\eta$. The dots indicate the initial configurations.}
\end{figure}
\addtocounter{example}{-3}	
\begin{example}[continued]\label{ex:coordination}
The single-population game with  binary action set and reward function \eqref{cost-12coordination} is a potential game with potential function $\Phi(x)=x_1^2+\gamma x_2^2$. It is not a contractive game since the set of Nash equilibria $\mc X^*$ is not convex. Figure \ref{fig:coordination} illustrates the first entry $x_1^{(\eta)}$ of the logit equilibria as a function of the noise level $\eta$ in the special cases $\gamma=1$ and $\gamma=2$, with continuous and dashed curves representing asymptotically stable logit equilibria and unstable logit equilibria, respectively. For low enough noise levels, we observe the existence of three logit equilibria that approach the Nash equilibria of the game in vanishing noise limit. Hence, in particular, $\mc X^{(0)} = \mc X^*$. Moreover, for low enough noise levels, the limit logit equilibria $\delta^{(1)}$ and $\delta^{(2)}$ are locally asymptotically stable, whereas the limit logit equilibrium $x^{*}$ is unstable. We remark that the asymptotic stability of the strict Nash equilibria $\delta^{(1)}$ and $\delta^{(2)}$ is not implied by the asymptotic stability of interior ESSs, since the strict equilibria are monomorphic. On the other hand, when the noise level $\eta$ is sufficiently large, a globally asymptotically stable logit equilibrium can be observed, where the population is equally split between the two available actions.

\end{example}
\begin{example}[continued]\label{ex:RSP}	
The single population game with ternary action set and reward function \eqref{cost-RSP} is never potential since 
%
%
	$$
	\ba{rcl}
	\ds\!\left(\!\frac{\partial}{\partial x_{1}}\!-\!\frac{\partial}{\partial x_{2}}\!\right)\left(r_{3} - r_{2}\right)\!\!\! 
	&=&\!\!\! \gamma - 3\\
	&\ne&\!\!\!	\gamma + 3\\
	&=&\!\!\!\ds 		\left(\!\frac{\partial}{\partial x_{3}}\!-\!\frac{\partial}{\partial x_{2}}\!\right)\left(r_{1} - r_{2}\right)\,,
	\ea$$
so that \eqref{eq:pot_games} is violated for every $\gamma$. 
However, for every two configurations $x$ and $y$ in $\mc X$, we have
$$
\sum_{i\in\mc S}(y_{i}-x_{i})(r_{i}(y)-r_{i}(x))
 =  \gamma||y-x||^2
 \ge  0 \,.
$$ 
Hence, \eqref{contractive} is satisfied (with $\alpha = \1$) and the population game is contractive if and only if $\gamma=0$. Therefore, in this case, for every noise level $\eta>0$, there exists a globally asymptotically stable logit equilibrium  $x^{(\eta)}$ that approaches $\mc X^*$ in the vanishing noise limit. The top-left panel of Figure \ref{fig:RSP} illustrates the solutions of the logit dynamic with noise level $\eta = 0.1$ for different initial conditions \tcb{when $\gamma=0$}. The other panels of Figure \ref{fig:RSP} show the solutions of the logit dynamics for different noise levels and initial configurations for the special case $\gamma=2$. In this case, the population game is not contractive and the dynamic exhibits a richer behavior. In particular, with noise level $\eta = 0.2$, the configuration converges to a strict Nash equilibrium of the game depending on the initial condition (except for the initial condition $x(0) = (1/3,1/3,1/3)$, which is an unstable logit equilibrium). When $\eta = 0.5$,  the presence of a locally attractive limit cycle is observed. Finally, when $\eta = 1$, the system converges to the globally asymptotically stable logit equilibrium $(1/3,1/3,1/3)$. \tcb{The local asymptotic stability of the strict equilibria of the game for small noise levels will be proved in Theorem \ref{thm:logit}, whereas the global asymptotic convergence of the dynamics to the center of the simplex when the noise level is large will be proved in Theorem \ref{thm:noise}.}
\end{example}
\begin{example}[continued]\label{ex:2pop2link}	
\begin{figure}
	\centering
	\includegraphics[width=6cm]{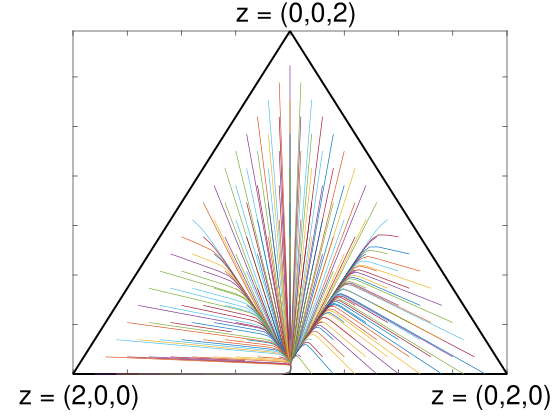}\\[2pt]
	\includegraphics[width=6cm]{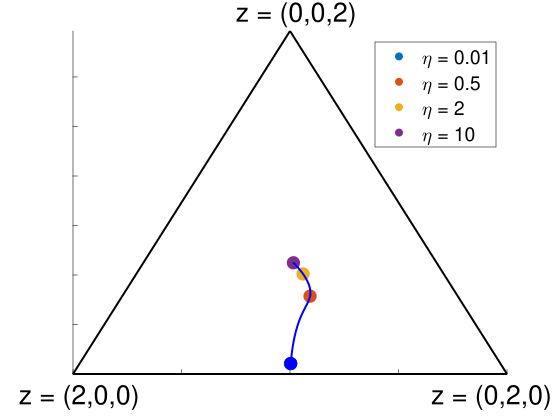}
	\caption{\label{fig:plot_cianfanelli}\emph{Top}: Solutions of the logit dynamics with noise level $\eta = 0.01$ in the game of Example \ref{ex:2pop2link} for different initial configurations, all converging to a globally asymptotically stable logit equilibrium. 
	\emph{Bottom}: The curve of logit equilibria as the noise level $\eta$ varies in the interval $[0.01,10]$. \tcb{In both the plots the variables $z_i(x) = \sum_{p \in \mc P}x_{ip}$ are reported.}}
\end{figure}
For the two-populations game with ternary action set and reward function as in \eqref{eq:reward_cianfanelli}, we have 
$$\ba{rcl}
\ds\!\left(\!\frac{\partial}{\partial x_{12}}\!-\!\frac{\partial}{\partial x_{31}}\!\right)\left(r_{11} - r_{21}\right)\!\!\! 
&=&\!\!\! 	-1\\
&\ne&\!\!\!	0\\
&=&\!\!\!\ds 		\left(\!\frac{\partial}{\partial x_{11}}\!-\!\frac{\partial}{\partial x_{21}}\!\right)\left(r_{12} - r_{31}\right)\,,
\ea$$
so that \eqref{eq:pot_games} is violated. Hence, this population game is not potential.  Moreover, consider the two configurations 
$$x=\left(\ba{cc}1&0\\0&1/3\\0&2/3\ea\right)\,,\qquad y=\left(\ba{cc}1/3&1\\2/3&0\\0&0\ea\right)\,,$$
whose associated reward matrices are
$$
r(x) = -\left(\ba{cc}1&2\\1&1/3\\2&1\ea\right)\,, \qquad r(y) = -\left(\ba{cc}4/3&2\\1&2/3\\2&1\ea\right)\,.
$$
Then,
$$y-x = \left(\!\ba{cc}\!-2/3\!&\!1\!\\\!2/3\!&\!-1/3\!\\\!0\!&\!-2/3\!\ea\!\right)\,,\quad \!\!\! r(y)-r(x)  = - \left(\ba{cc}1/3&0\\0&1/3\\0&0\ea\right)\,.
$$
so that
$$\ds\sum_{p\in\mc P} \sum_{i\in\mc S}\alpha_p(y_{ip}-x_{ip})(r_{ip}(y)-r_{ip}(x))
 =  \ds\frac{2\alpha_1 + \alpha_2}9>0\,.$$
Therefore, condition  \eqref{contractive} is violated for every $\alpha_1, \alpha_2 > 0$. Hence, this population game is not weighted-contractive.
Nevertheless, numerical simulations, such as the one reported in the top panel of Figure \ref{fig:plot_cianfanelli} for the special case of noise level $\eta=0.01$, suggest that, for every noise level $\eta>0$, this population game admits a globally asymptotically stable logit equilibrium $x^{(\eta)}$. 
In fact, as shown in the bottom panel of Figure \ref{fig:plot_cianfanelli}, such globally asymptotically stable logit equilibria $(x^{(\eta)})_{\eta > 0}$ appear to form a curve connecting the unique Nash equilibrium $x^*$ of the game (reached in the vanishing noise limit) to the configuration where each population is equally split among the available actions (reached in the limit as the noise level grows large). \tcb{The existence of such a curve of globally asymptotically stable equilibria will be explained by Theorem \ref{thm:conv_global}.}
\end{example}

\section{Main Results}\label{sec:results}

In this section, we present our main results. 
Section \ref{sec:arb_graphs} presents our results concerning the stability of the logit dynamics in population games for low enough noise levels. Section \ref{sec:high_noise} characterizes the dynamics in the large noise limit. Then, Section \ref{sec:parallel} provides sufficient conditions for global asymptotic stability of logit equilibria for every noise level.

\subsection{Asymptotic stability of strict Nash equilibria for low noise}
\label{sec:arb_graphs}
Our first result concerns the asymptotic stability of strict Nash equilibria under the logit dynamics for low enough noise levels. 

Let $\wz$ in $\mc{X}^*_s$ be a strict Nash equilibrium of a population game \eqref{def:pop-game}. For every $p$ in $\mc P$, let $s_p$ denote the action played by population $p$ in $x^*$, so that $x^*_{s_pp} = v_p$ for every $p$ in $\mc{P}$. For $\epsilon$ in $[0,1]$, let
\be\label{eq:o_eps}
\mc O_\epsilon(\wz)= \{x \in \mc{X}:  x_{s_p p} \ge v_p(1-\epsilon)\,,\ \forall p \in \mc{P}\}
\ee
denote the set of configurations such that at least a fraction $1-\epsilon$ of players of each population $p$ uses action $s_p$. Then, we have the following technical result that will prove instrumental in deraving our main result. 
\begin{lemma}\label{lemma:G}
Let $x^*$ in $\mc X^*_s$ be a strict Nash equilibrium of a population game \eqref{def:pop-game}. Let $F: \mc X \times (0,+\infty) \to \mc X$ be defined as in \eqref{def:G}. Then, there exists $\bar \epsilon > 0$ such that:
	\begin{enumerate}
		\item[(i)] for every $\epsilon$ in $(0,\bar \epsilon]$ there exists $\eta_\epsilon > 0$ such that \tcb{$F(x,\eta)$ belongs to $\mc O_\epsilon(\wz)$ for every $x$ in $\mc O_{\epsilon}(x^*)$ and $\eta$ in $(0,\eta_\epsilon]$};
		\item[(ii)]
		the function $\bar F: \mc O_{\bar\epsilon}(\wz) \times \R_+ \to \mc X$ defined by
		\be\label{barF}
		\bar{F}(x,\eta) = 
		\begin{cases}
			F(x,\eta) \quad & \text{if} \ \eta>0\,, \\
			\wz & \text{if} \  \eta=0\,,
		\end{cases}
		\ee
		is continuously differentiable and
		\begin{equation}
			\label{eq:J}
			\lim_{\eta \to 0^+}\frac{\partial F_{jq}}{\partial x_{ip}}(x,\eta) ={0} \,,
		\end{equation}
		for every two populations $p$ and $q$ in $\mc P$, actions $i$ in $\mc S_p$ and $j$ in $\mc S_q$, and configuration $x$ in $\mc O_{\bar\epsilon}(\wz)$.
	\end{enumerate} 
\end{lemma}
\begin{IEEEproof} See Appendix \ref{app:G}.
\end{IEEEproof}\medskip

We are now in a position to state and prove the following result on the asymptotic stability of strict Nash equilibria of every population game under the logit dynamics for low enough noise levels. 

\begin{theorem}
	\label{thm:logit} Consider a population game \eqref{def:pop-game}. 
	For every strict Nash equilibrium $\wz$ in $\mc{X}_s^*$, there exists $\tilde{\eta}>0$ and a differentiable curve $(x^{(\eta)})_{0\le \eta < \tilde\eta}\subseteq \mc X$ such that:
	\begin{enumerate} 
	\item[(i)] $x^{(\eta)}\in\mc X^{(\eta)}$ is an asymptotically stable logit equilibrium for every $\eta$ in $(0,\tilde\eta)$; 
	\item[(ii)] $x^*=x^{(0)}\in\mc X^{(0)}$ is a limit logit equilibrium. 
	\end{enumerate}
Hence, in particular, $\mc X _s^* \subseteq \mc X^{(0)}$.
\end{theorem}

	\begin{IEEEproof}
Lemma \ref{lemma:G} ensures the existence of a small $\eps>0$ such that the function $\bar{F} : \mc O_{\epsilon}(\wz) \times \R_+ \to \mc{X}$ defined by \eqref{barF}
is continuously differentiable and 
\tcb{$\bar F(x,\eta)$ belongs to $\mc O_{\epsilon}(\wz)$ for every $x$ in $\mc O_{\epsilon}(x^*)$ and $\eta$ in $[0,\eta_{\epsilon}]$. This implies by Brouwer's fixed point theorem that $x \mapsto \bar F(x,\eta)$ admits at least a fixed point $x^{(\eta)}$ in $\mc O_{\epsilon}(x^*)$ for every $\eta$ in $[0,\eta_\eps]$. Now, let $f : \mc O_\epsilon(x^*) \times \R_+ \to \mc X$ defined by $$f(x,\eta) = \bar{F}(x,\eta)-x\,.$$ Now, notice that $\mc X^{(\eta)} \cap \mc O_{\epsilon(x^*)}$ coincides with the the zeros of $x \mapsto f(x,\eta)$ for every $\eta$ in $(0,\eta_\eps]$ and that ${f}(\wz,0)=0$. Hence, the fact that $f$ is continuously differentiable, \eqref{eq:J},
and the Implicit Function Theorem \cite[Theorem 9.18]{rudin1976principles} imply the existence of a differentiable curve of $(x^{(\eta)})_{0\le\eta< \tilde\eta}$ in $\mc X^{(\eta)}\cap\mc O_{\eps}(\wz)$ such that $x^{(0)}=\wz$.}
Furthermore, 
\eqref{eq:J} implies asymptotic stability of the logit equilibria $x^{(\eta)}$ for small enough noise levels $\eta>0$.
\end{IEEEproof}\medskip

\subsection{Contractivity and global exponential stability of the logit dynamics for large noise levels}\label{sec:high_noise}
Our second main result concerns the global asymptotic stability of logit equilibria for large noise levels. 
Its proof builds on contractive systems theory: in particular, we shall use the following result establishing a sufficient condition for a dynamical system to be contractive in the $l_1$-distance.
\begin{lemma} 
	\label{prp:contractivity}
Consider the continuous-time dynamical system  $\dot x=f(x)$ for a continuously differentiable vector field $f:\mathds{R}^n \to \mathds{R}^n$.  Let $\mc{X} \subseteq \mathds{R}^n$ be positively invariant set  and assume that there exists a constant $\alpha>0$ such that
\be\label{eq:diag_dominant}
\max_{j} \left\{\frac{\partial f_j}{\partial x_j}(x)+\sum_{i\neq j} \left|\frac{\partial f_i}{\partial x_j}(x)\right|\right\} \le -\alpha\,,\quad \forall x\in\mc X\,.
\ee
Let $x(t)$ and $\tilde x(t)$ denote the solutions of the dynamical system at time $t$ corresponding to initial conditions $x(0)$ and $\tx(0)$ in $\mc{X}$, respectively. Then:
	\begin{enumerate}
		\item[(i)] for every $t \ge 0$
		\begin{equation}
			\label{eq:l1_contract}
			\tcb{||x(t) - \tx(t)|| \le ||x(0)-\tx(0)||} e^{-\alpha t};
		\end{equation}
		\item[(ii)] there exists a globally exponentially stable equilibrium point $x^*$ in $\mc{X}$.
	\end{enumerate}
\end{lemma} 
\begin{IEEEproof}{See Appendix \ref{sec:proof-prp-contractivity}}. \end{IEEEproof}\medskip

\begin{theorem}
	\label{thm:noise}
		For every population game \eqref{def:pop-game}, there exists \tcb{$\bar\eta>~0$} such that \eqref{eq:logit} admits a globally exponentially stable logit equilibrium $x^{(\eta)}$ for every noise level $\eta \ge \bar\eta$.
		Moreover,
			$$
			\lim_{\eta \to +\infty} x^{(\eta)}_{ip} = \1_{\mc S_p}(i) \frac{v_p}{|\mc S_p|}\,, 
			$$
			for every action $i$ in $\mc S$ and population  $p$ in $\mc P$. 
\end{theorem}
	\begin{IEEEproof}
Let $$\Delta_{ij}^p(x) = r_{ip}(x)-r_{jp}(x)$$ be the difference between the reward of action $i$ and that of action $j$ for players of population $p$ in configuration $x$. \tcb{Since $x \mapsto F(x,\eta)$ is continuously differentiable (c.f. Lemma \ref{lemma0}(iv)), we can take the partial derivative of both sides of \eqref{def:G}, and get that}
\be\label{eq:Jf}
	\frac{\partial F_{ip} (x,\eta)}{\partial x_{jq}}=
	\frac{\ds v_pe^{r_{ip}(x)/\eta}\sum\nolimits_{k\ne i}
	\frac{\ds\partial \Delta_{ik}^p(x)}{\ds\partial x_{jq}} e^{r_{kp}(x)/\eta}}{\ds\eta\left(\sum\nolimits_{k} e^{r_{kp}(x)/\eta}\right)^2}\stackrel{\eta \to +\infty}{\longrightarrow}0\,,
\ee
for every two populations $p$ and $q$ in $\mc P$, actions $i$ in $\mc S_p$ and $j$ in $\mc S_q$, and configuration $x$ in $\mc{X}$.
 Now, for every configuration $x$ in $\mc X$ and noise level $\eta>0$, let $$f(x,\eta) = F(x,\eta)-x\,,$$ so that \eqref{eq:logit} reads
 $\dot x = f(x,\eta)$, \tcb{with $f$ continuously differentiable}. 
It then follows from \eqref{eq:Jf} that 
$$
\!	\lim_{\eta \to +\infty} \max_{p\in\mc P,j\in\mc S_p} \left\{\frac{\partial f_{jp}}{\partial x_{jp}}(x,\eta)+
	\!\!\!\sum_{(i,q)\neq (j,p)} \left|\frac{\partial f_{iq}}{\partial x_{jp}}(x,\eta)\right|\right\}  = -1\,, $$
	for every $x$ in $\mc{X}$.
Hence, there exists $\bar\eta \tcb{>} 0$ such that,  
\eqref{eq:diag_dominant} is satisfied with $\alpha=1/2$ by $f(x,\eta)$ for every  $\eta\ge\bar\eta$. 
Hence, Lemma \ref{prp:contractivity} implies  that there exists a globally exponentially stable logit equilibrium $x^{(\eta)}$ for every noise level $\eta\ge\bar\eta$, thus proving the first part of the statement. %
From \tcb{\eqref{eq:logit}}, we have that 
	$$
\lim_{\eta \to +\infty} x^{(\eta)}_{ip} = \lim_{\eta \to +\infty} F_{ip}(x^{(\eta)},\eta)\,,
	$$
	for every action $i$ in $\mc S$ and population  $p$ in $\mc P$. 
Hence, \tcb{Lemma \ref{lemma0}(ii)} implies that
$$
\lim_{\eta \to +\infty} x^{(\eta)}_{ip} = \1_{\mc S_p}(i) \frac{v_p}{|\mc S_p|}\,,
$$
for every action $i$ in $\mc S$ and population  $p$ in $\mc P$.  
\end{IEEEproof}\smallskip
	
	The behavior of the logit dynamics in the large noise limit established in Theorem \ref{thm:noise} is well illustrated in Figures \ref{fig:coordination}-\ref{fig:plot_cianfanelli} for the population games of Examples \ref{ex:coordination}-\ref{ex:2pop2link}.
	Notice that, for population games that admit multiple strict equilibria, on the one hand Theorem \ref{thm:logit} implies that the strict equilibria are asymptotically stable under the logit dynamics for low enough noise levels, on the other hand, as the noise grows large, Theorem \ref{thm:noise} implies that the dynamics admit a globally asymptotically stable logit equilibrium. These results together suggest that, for population games that admit multiple strict equilibria, the logit dynamics exhibit a bifurcation, as illustrated in the potential game of Example \ref{ex:coordination}. The emergence of bifurcations will be also shown in the next section in population games that do not admit a potential function (c.f. Example \ref{ex:toso}).
 \begin{remark}
 Analogous arguments based on $l_1$-contractivity (or non-expansiveness) of the system have been recently used in different applicative contexts such as dynamical flow networks \cite{lovisari2014stability,Como.Lovisari.Savla:2015,como2017resilient} and systems biology \cite{Margaliot:12} 
 \end{remark}\smallskip	
 \begin{remark}
 	An alternative proof of Theorem \ref{thm:noise} leverages the theory of passive systems. In fact, the logit dynamics with noise level $\eta$ are known to be $\delta$-passive with surplus $\eta$ \cite{park2018passivity}. Moreover, population games with $M$-Lipschitz reward functions are $\delta$-antipassive with deficit $M$. Hence, one can prove by \cite[Theorem 5]{park2019population} that if $\eta>M$ the logit dynamics admit a globally asymptotically stable logit equilibrium.
 \end{remark}

\subsection{Global asymptotic stability of logit equilibria for monotone separable population games}
\label{sec:parallel}
In this section, we prove that a certain class of population games, as characterized by the following definition, admits a globally asymptotically stable logit equilibrium $x^{(\eta)}$ for every noise level $\eta>0$.

\begin{definition}\label{def:sep}
	A population game \eqref{def:pop-game} is \emph{monotone separable} if the action sets of all populations $p$ are a direct product
$$\mc S_p = \mc S = \prod_{k = 1}^{l} \mc A_{k}\,,$$
	and there exists a non-negative vector of weights $\theta$ in $\R_{+}^\mc P$ such that the reward function admits the decomposition
	\be\label{eq:cost}
	r_{ip}(x) = \sum_{k=1}^l \varphi^{(k)}_{i_kp}(z^{(k)}_{i_k}(x))\,, \qquad\forall x\in\mc X\,,\ee
for every $p$ in $\mc P$ and $i$ in $\mc S_p$,
where 
$\varphi^{(k)}_{ap}:\R_+\to\R\,,$ is a non-increasing function of the aggregate weighted mass 
	\be\label{eq:z}
	z^{(k)}_{a}(x) = \sum_{p \in \mc P}\theta_p \sum_{i \in \mc S: i_k = a} x_{ip}\,,
	\ee
	of players playing actions with $k$-th entry equal to $a$, 
	for every population $p$ in $\mc P$, $k = 1,\cdots,l$,  and $a$ in $\mc A_{k}$. 
\end{definition}


\begin{theorem}\label{thm:conv_global}
	Consider a monotone separable population game \eqref{def:pop-game}. Then: 
	\begin{enumerate}
		\item[(i)] for every noise level $\eta>0$ there exists a globally asymptotically stable logit equilibrium $x^{(\eta)}$;
		\item[(ii)] $\eta \mapsto x^{(\eta)}$ is a \tcb{continuously differentiable curve on $(0,+\infty)$} such that
		\begin{equation}\label{eq:thm3}
			x^{(0)} := \lim_{\eta\to 0^+} x^{(\eta)} \in \mc X^{(0)}\,,
		\end{equation}
is a limit logit equilibrium and
\be\label{eq:x_inf}
\lim_{\eta \to +\infty} x^{(\eta)}_{ip} = \frac{v_p}{|\mc S|}\,, 
\ee
for every action $i$ in $\mc S$ and population  $p$ in $\mc P$. 
	\end{enumerate}
\end{theorem}
\begin{IEEEproof}
(i) For every $k = 1,\cdots,l$, let $$\mc Z^{(k)}=\Bigg\{z^{(k)} \in \R_+^{\mc{A}_k}: \sum_{a\in\mc A_k}z_a^{(k)} = \sum_{p\in\mc P} \tcb{\theta_p} v_p\Bigg\}\,,$$ and, for every $a$ in $\mc A_{k}$ and $z^{(k)}$ in $\mc Z^{(k)}$,
$$f^{(k)}_a(z^{(k)},\eta) = \sum_{p\in\mc P} \theta_p v_p \frac{\ds \exp(\varphi^{(k)}_{ap}(z_a^{(k)})/\eta)}{\ds\sum_{s\in\mc A_{k}}\exp(\varphi^{(k)}_{sp}(z_s^{(k)})/\eta)}-z_a^{(k)}\,,$$
Now, let
$$\psi^{(k)}_{ap}=\varphi^{(k)}_{ap}(z_a^{(k)})/\eta\,,\qquad 1\le k\le l\,.$$  
If $x$ is a solution of the logit dynamic \eqref{eq:logit}, then
$$
\ba{rcl}\dot z_a^{(k)}
&=&\ds\sum_{p\in\mc P}\theta_p\sum_{\substack{i \in \mc S:\\ i_k = a}} \dot x_{ip}
=\ds\sum_{p\in\mc P}\theta_p\sum_{\substack{i \in \mc S:\\ i_k = a}} \left(F_{ip}(x,\eta)-x_{ip}\right)
\\[25pt]
&=&\ds\sum_{p\in\mc P}\theta_p \sum_{\substack{i \in \mc S:\\ i_k = a}} \frac{v_p\exp(r_{ip}(x)/\eta)}{\ds\sum_{j \in \mathcal{S}} \exp(r_{jp}(x)/\eta)}-\sum_{p \in \mc P} \theta_p \sum_{\substack{i \in \mc S:\\ i_k = a}}x_{ip}\\[25pt]
&=&\ds\sum_{p\in\mc P}\theta_p  \frac{\ds v_p\sum_{i: i_k = a} \exp\Big(\sum_{1\le h \le l } \psi^{(h)}_{i_hp}\Big)}{\ds\sum_{j\in\mc S}\exp\Big(\sum_{1\le h \le l }  \psi^{(h)}_{j_hp}\Big)}-z_a^{(k)}\\[25pt]
&=&\ds\sum_{p\in\mc P}\theta_p  \frac{\ds v_p\sum_{i: i_k = a} \prod_{1\le h \le l }\exp\left( \psi^{(h)}_{i_hp}\right)}
{\ds\sum_{j \in \mc S} \prod_{1\le h \le l } \exp\left(\psi^{(h)}_{j_hp}\right)} - z_a^{(k)}
\\[25pt]
&=&\ds\sum_{p\in\mc P}\theta_p \frac{v_p \ds\exp(\psi^{(k)}_{ap})\prod_{h \neq k}\sum_{s \in \mc A_{h}}\exp(\psi^{(h)}_{sp})}
{\ds\prod_{1\le h\le l}\sum_{s \in \mc A_{h}}\exp(\psi^{(h)}_{sp})} - z_a^{(k)}\\[25pt]
&=&\ds\sum_{p\in\mc P}\theta_p \frac{\ds v_p\exp(\psi^{(k)}_{ap})}{\ds\sum_{s \in \mc A_{k}}\exp(\psi^{(k)}_{sp})} - z_a^{(k)} \\[25pt]
	&=&\ds f_a^{(k)}(z^{(k)},\eta)\,,
\ea$$
for every $k = 1,\cdots,l$, $a$ in $\mc A_{k}$, and $z^{(k)}$ in $\mc Z^{(k)}$. This proves that $z^{(k)}$  is a solution of the autonomous dynamical system \be\label{dyn-sys-k}\dot z^{(k)}=f^{(k)}(z^{(k)},\eta)\,,\qquad k = 1,\cdots,l\,.\ee Now, observe that for every $i,j$ in $\mc A_{k}$ with $j\ne i$ and $z^{(k)}$ in $\mc Z^{(k)}$,
\be\label{eq:metzler}
\ds\frac{\partial f^{(k)}_i}{\partial z^{(k)}_j}
=-\sum_{p\in\mc P}\frac{\theta_p v_p\exp(\psi^{(k)}_{ip}+\psi^{(k)}_{jp})}{\eta\left(\sum_{s\in\mc A_{k}}\exp(\psi^{(k)}_{sp})\right)^2}\left(\varphi^{(k)}_{jp}\right)'
\ge0\,,
\ee
where the inequality follows since $\varphi_{jp}^{(k)}$ is non-increasing by Definition \ref{def:sep}. Moreover, $$\sum_{i \in \mc A_{k}} f^{(k)}_i(z^{(k)},\eta)=\sum_{p\in\mc P}\theta_pv_p-\sum_{i \in \mc A_{k}} z^{(k)}_i,$$ for every $z^{(k)}$ in $\mc Z^{(k)}$, so that 
\be\label{eq:diag_dom}\sum_{i\in\mc A_k}\frac{\partial f_i^{(k)}}{\partial z^{(k)}_j}=-1\,.\ee  Therefore, 
$$\max_{j\in\mc A_k} \left\{\frac{\partial f^{(k)}_j}{\partial z_j^{(k)}}+\sum_{i\neq j} \left|\frac{\partial f^{(k)}_i}{\partial z_j^{(k)}}\right|\right\}
=\max_{j\in\mc A_k} \sum_{i\in\mc A_k}\frac{\partial f^{(k)}_i}{\partial z^{(k)}_j}
=-1\,,$$
where the first equality follows from \eqref{eq:metzler} and the second one from \eqref{eq:diag_dom}.
\tcb{Moreover, since $\mc X$ is positively invariant for $x(t)$ (c.f. Lemma \ref{lemma:existence-uniqueness}(i)), then $\mc Z^{(k)}$ is positively invariant for $z^{(k)}(t)$ for every $k = 1,\cdots,l$, as
$$
\sum_{a \in \mc A_k} z^{(k)}_a(t) = \sum_{p \in \mc P} \theta_p \sum_{a \in \mc A_k} \sum_{i \in \mc S: i_k = a} x_{ip}(t) = \sum_{p \in \mc P} \theta_p v_p\,.
$$}
It then follows from Lemma \ref{prp:contractivity} that the dynamical system \eqref{dyn-sys-k}
admits a globally exponentially stable equilibrium point $z^{(k,\eta)}$ in $\mc Z^{(k)}$. 
Observe that, since $$z^{(k)}(t)\stackrel{t\to+\infty}{\longrightarrow}z^{(k,\eta)}\,,$$ for every $k = 1,\cdots,l$, then  
$$
F_{ip}(x(t),\eta)
=\frac{v_p\exp(\sum_{k = 1}^l\varphi^{(k)}_{i_kp}(z^{(k)}_{i_k}(t))/\eta)}{\sum_{j\in\mc S}\exp(\sum_{k = 1}^l\varphi^{(k)}_{j_kp}(z_{j_k}^{(k)}(t))/\eta)}
\stackrel{t\to+\infty}{\longrightarrow}x^{(\eta)}_{ip}\,,$$
for every $p$ in $\mc P$ and $i$ in $\mc S$,  where \be\label{eq:x*}x^{(\eta)}_{ip}=\frac{v_p\exp(\sum_{k = 1}^l\varphi^{(k)}_{i_kp}(z^{(k,\eta)}_{i_k}))/\eta)}{\sum_{j\in\mc S}\exp(\sum_{k = 1}^l\varphi^{(k)}_{j_kp}((z_{j_k}^{(k,\eta)})/\eta)}\,.\ee 
Finally, since $\dot x=F(x,\eta)-x$, the fact that $$F(x(t),\eta)\stackrel{t\to+\infty}{\longrightarrow}x^{(\eta)}\,,$$ for every initial condition $x(0)$
implies that 
$$x(t)\stackrel{t\to+\infty}{\longrightarrow}x^{(\eta)}\,,$$
thus proving point (i) of the claim.

(ii)
Equations \eqref{eq:metzler} and \eqref{eq:diag_dom} imply by the Gershgorin Circle Theorem that the Jacobian matrix $\nabla_z f^{(k)}(z,\eta)$ is invertible for every $z$ in $\mc Z^{(k)}$ and $\eta>0$. Since $f^{(k)}(z,\eta)$ admits a unique zero in $\mc Z^{(k)}$ for every $\eta>0$ by point (i), hence \tcb{the continuous differentiability of $f^{(k)}$ (which follows from that of $F$, c.f. Lemma \ref{lemma0}(iv)-(v)) and} the Implicit Function Theorem ensure the existence of a $\mc C^1$ map $\eta \mapsto z^{(k,\eta)}$ \cite[Theorem 9.18]{rudin1976principles} for every $k = 1,\cdots,l$. Equation \eqref{eq:x*} and the differentiability of the reward function then imply the existence of a continuously differentiable curve $\eta \mapsto x^{(\eta)}$. Finally, \eqref{eq:thm3} follows from \tcb{the definition of limit logit equilibrium} and \eqref{eq:x_inf} follows from Theorem \ref{thm:noise}.
\end{IEEEproof}\medskip
\addtocounter{example}{-1}	
\begin{example}[continued]
	The population game is monotone separable with respect to the decomposition $
	\mc S = \mc A_{1} = \{1,2,3\}$ with $\theta = \ones$. Indeed, by \eqref{eq:z},
	$$
	z_a^{(1)}(x) = x_{a1} + x_{a2}\,, \quad a = 1,2,3\,.
	$$
	Moreover, by letting
	$$
	\ba{lcl}\varphi^{(1)}_{11}(z^{(1)}_1)=-z_1^{(1)}\,,&\ & \varphi^{(1)}_{12}(z^{(1)}_1)=-2\,,\\[7pt]
	\varphi^{(1)}_{21}(z^{(1)}_2)=-1\,,&\ & \varphi^{(1)}_{22}(z^{(1)}_1)=-z_2^{(1)}\,,\\[7pt]
	\varphi^{(1)}_{31}(z^{(1)}_{3})=-2\,,&\ & \varphi^{(1)}_{32}(z^{(1)}_3) = -1. \ea
	$$
	the reward function \eqref{eq:reward_cianfanelli} is in the form \eqref{eq:cost}. Hence, Theorem \ref{thm:conv_global} establishes the existence of a continuously differential curve of globally asymptotically stable logit equilibria, as illustrated in Figure~\ref{fig:plot_cianfanelli}.
\end{example}

	We remark that in general monotone separable games are neither (weighted) contractive nor potential, as Example \ref{ex:2pop2link} proves.
	Hence, the global asymptotic stability of the logit equilibria in this class of population games is a novel result. 

\section{Application to heterogeneous routing games}\label{sec:routing}
Monotone separable population games particularly arise in the context of routing games, as explained below. 

		Consider a finite directed multigraph $\mathcal{G}=(\mathcal{N},\mathcal{E})$ with set of nodes $\mc N$ and set of links $\mc E$. 
		For two nodes $o\ne d$ in $\mc N$, a route from $o$ to $d$ is a $k$-tuple of links $i=(i_1,\ldots,i_k)$ such that the tail node of $i_1$ is $o$, the head node of $i_k$ is $d$, the head node of $i_{h-1}$ coincides with the tail node of $i_h$ for all $h = 2,\cdots,k$, and no node is visited twice. 
Node $d$ is said to be reachable from node $o$ in $\mc G$ if there exists at least one route from $o$ to $d$. 

For a pair of distinct nodes $o\ne d$ in $\mc V$, such that $d$ is reachable from $o$, let $\mc S$ denote the set of routes from $o$ to $d$ and $L$ in $\{0,1\}^{\mc{E} \times \mc{S}}$ denote the link-route incidence matrix, with entries $L_{ei}=1$ if link $e$ belongs to route $i$, and $L_{ei}=0$ otherwise.
A multigraph is called \emph{parallel} if every link belongs to at most one route, i.e., $L\ones \le \ones$.

	Let $\mc E_p \subseteq \mc E$ be the set of links available to population $p$, and $\mc S_p \subseteq \mc S$ the set of paths $i$ such that $i_h$ belongs to $\mc E_p$ for every $h = 1,\cdots, k$.
	 A \emph{routing game} is a population game \eqref{def:pop-game} with action sets $\mc S_p \subseteq \mc S$ and reward function 
	 \be\label{eq:reward_routing}
	 \ba{rcl}
	 r_{ip}(x) & = &\ds -\sum_{e \in \mc E} L_{ei} \tau_{ep}(y_e(x))\,,\\[15pt] \tcb{y_e(x)} & \tcb{=} &\ds \tcb{\sum_{p \in \mc P}\theta_p \sum_{i \in \mc S} L_{ei} x_{ip }}\,,
	 \ea
	 \ee
	 where, for every population $p$ in $\mc P$ and link $e$ in $\mc E$,
	  $$\tau_{ep}: \R_+ \to \R_+\,,$$ is a non-decreasing function returning the cost perceived by players in population $p$ for using link $e$ as a function of 
	the weighted mass $y_e(x)$ of players using link $e$.

Routing games with $|\mc P|=1$ are called homogeneous. As shown in \cite{beckmann1956studies}, homogeneous routing games are potential games with concave potential function
$$
\Phi(x) = - \sum_{e \in \mc E} \int_0^{\sum_{i \in \mc S} \tcb{L}_{ei}x_i} \tau_e(s) \de s\,,
$$ 
hence the logit dynamics admit a globally asymptotically stable logit equilibrium that approaches the convex set of Nash equilibria of the game in the vanishing noise limit. On the other hand, it is known that heterogeneous routing games are potential games only under very restrictive conditions on the link cost functions, such as the one considered in \cite[Lemma 4.2]{farokhi2013heterogeneous}. In fact, Example \ref{ex:2pop2link} may be interpreted as a non-potential heterogeneous routing game with $|\mc P|=2$ on a parallel multigraph with three links. The next example illustrates the behavior of the logit dynamics in a non-potential heterogeneous routing game on more complex topologies.
\begin{figure}
	\centering
	\centering
	\begin{tikzpicture}[scale=1, transform shape]
		\node[draw, circle] (1) at (0,0)  {o};
		\node[draw, circle] (2) at (2,1)  {a};
		\node[draw, circle] (3) at (2,-1)  {b};
		\node[draw, circle] (4) at (4,0) {d};
		
		\path (1) edge [->] node [above] {$e_1$} (2);
		\path (2) edge [->, bend left = 20] node [above] {$e_2$} (4);
		\path (2) edge [->, bend right = 20] node [above] {$e_3$} (4);
		\path (1) edge [->] node [above] {$e_4$} (3);
		\path (3) edge [->, bend right = 20] node [above] {$e_6$} (4);
		\path (3) edge [->, bend left = 20] node [above] {$e_5$} (4);
	\end{tikzpicture}\\[5pt]
	\resizebox{\columnwidth}{!}{
		\begin{tabular}{c|c|c|c|c|c|c|c}
			$p$ & $v_p$ & $\tau_{1p}(y_1)$ & $\tau_{2p}(y_2)$ & $\tau_{3p}(y_3)$ & $\tau_{4p}(y_4)$ & $\tau_{5p}(y_5)$ & $\tau_{6p}(y_6)$ \\
			\hline
			1 & 1.4 & $y_1$ & $y_2$ &  & $y_4$ &  & $y_6$ \\ 
			2 & 1 & $y_1$ & 20$y_2$ & & $y_4$ & $y_5$+21 & \\ 
			3 & 1 & $y_1$ & & $y_3$+21 & $y_4$ & & 20$y_6$ \\ 
	\end{tabular}}\\[5pt]
	\includegraphics[width=9cm]{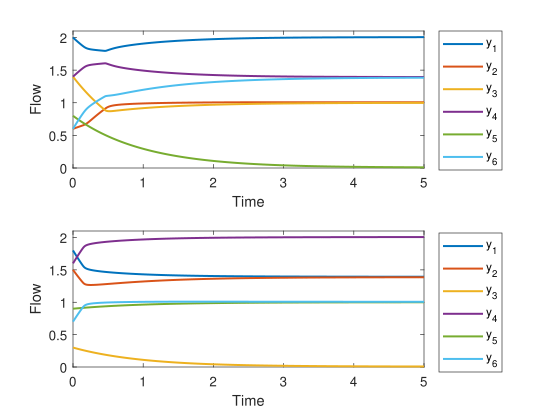}
	\caption{\label{fig:toso} The multigraph and the link cost functions of the routing game in Example \ref{ex:toso}. The bottom panels report two solutions of the logit dynamic with noise level $\eta = 0.05$ corresponding to different initial configurations. The curves represent the entries of the link flow $\tcb{y(x(t))}$.}
\end{figure}
\begin{example}\label{ex:toso}
	Consider the heterogeneous routing game on the multigraph $\mc G = (\mc N,\mc E)$ in the top panel of Figure \ref{fig:toso}, with link-route incidence matrix
		$$
		L = \left(\begin{matrix}
			1 & 1 & 0 & 0 \\
			1 & 0 & 0 & 0 \\
			0 & 1 & 0 & 0 \\
			0 & 0 & 1 & 1 \\
			0 & 0 & 1 & 0 \\
			0 & 0 & 0 & 1
		\end{matrix}\right).
		$$
		Let $|\mc P| = 3$, $\theta = \ones$, and the link cost functions as defined in Figure \ref{fig:toso}, where a blank space in $\tau_{ep}$ means that $e$ does not belong to $\mc E_p$.
		The game admits
		two strict Nash equilibria $\bar x, \tilde x$
		$$\bar x = \left(\begin{matrix}
			1.4 & 0 & 0 \\ 0 & 0 & 0 \\ 0 & 1 & 0 \\ 0 & 0 & 1\end{matrix}\right)\,, \quad \tilde x = \left(\begin{matrix}
			0 & 1 & 0 \\ 0 & 0 & 1 \\ 0 & 0 & 0 \\ 1.4 & 0 & 0 
		\end{matrix}\right)\,, 
		$$
		with corresponding link flow equal to
		$$\bar y = (1.4, 1.4, 0, 2, 1, 1), \quad \tilde y = (2, 1, 1, 1.4, 0, 1.4)\,,$$
		and a third (non-strict) Nash equilibrium
		$$
		\hat x =
		\left(\begin{matrix}
			357 & 160 & 0 \\ 0 & 0 & 160 \\ 0 & 260 & 0 \\ 357 & 0 & 260 
		\end{matrix}\right)/420\,. 
		$$
		Figure \ref{fig:toso} illustrates two solutions of the logit dynamic with noise level $\eta = 0.05$ corresponding to different initial conditions. In particular, the numerical simulations show that the dynamic does not admit a globally asymptotically stable logit equilibrium. In fact, the link flow converges to $\bar y$ or $\tilde y$ depending on the initial condition. Hence, strict Nash equilibria are asymptotically stable for low enough noise levels, as established by Theorem \ref{thm:logit}. In contrast, the non-strict equilibrium $\hat x$ is unstable. On the other hand, for sufficiently large noise levels, the numerical simulations show the existence of globally asymptotically stable logit equilibrium, consistently with Theorem \ref{thm:noise}.
\end{example}

The previous example shows that, in contrast with the homogeneous case, heterogeneous routing games may not admit globally asymptotically stable logit equilibria. In the remainder of the section, we provide sufficient conditions on the network topology for the existence of a globally asymptotically stable logit equilibrium that follow from Theorem \ref{thm:conv_global}.
\begin{definition}\label{def:series}
Consider $l$ multigraphs $\{\mc G^{(k)} = (\mc N^{(k)},\mc E^{(k)})\}_{k=1}^l$ and $l$ pairs of distinct nodes $o^{(k)} \neq d^{(k)}$ in $\mc N^{(k)}$ such that: 
	\begin{itemize}
		\item $o^{(k+1)} = d^{(k)}$ for every $k = l,\cdots,l-1$;
		\item $\mc N^{(k)} \cap \mc N^{(k+1)} = \{d^{(k)}\} = \{o^{(k+1)}\}$ for every $k = 1,\cdots,l-1$;
		\item $\mc N^{(k)} \cap \mc N^{(j)} = \emptyset$ if $|k-j| > 1$;
		\item $\mc E^{(k)} \cap \mc E^{(j)} = \emptyset$ for every $k \neq j$.
	\end{itemize}
Then, the \emph{series composition} of $\{\mc G^{(k)}\}_{k=1}^{l}$ is the multigraph $\mc G = (\mc N,\mc E)$ with node set $\mc N = \bigcup_{k=1}^l \mc N^{(k)}$ and link set $\mc E = \bigcup_{k=1}^l \mc E^{(k)}$.
\end{definition}
\begin{proposition}\label{prop:graph}
Routing games on series compositions of parallel multigraphs with $\mc S_p = \mc S$ for all $p$ in $\mc P$ are monotone separable.
\end{proposition}
\begin{IEEEproof}
See Appendix \ref{app:graph}.
\end{IEEEproof}
\begin{corollary}\label{cor:routing}
	Consider a routing game on a series composition of parallel multigraphs with $\mc S_p = \mc S$ for all $p$ in $\mc P$. Then: 
		\begin{enumerate}
			\item[(i)] for every noise level $\eta>0$ there exists a globally asymptotically stable logit equilibrium $x^{(\eta)}$;
			\item[(ii)] $\eta \mapsto x^{(\eta)}$ is a \tcb{continuously differentiable curve on $(0,+\infty)$} such that
			\begin{equation}
				x^{(0)} := \lim_{\eta\to 0^+} x^{(\eta)} \in \mc X^{(0)}\,,
			\end{equation}
			is a limit logit equilibrium and
			\be
			\lim_{\eta \to +\infty} x^{(\eta)}_{ip} = \frac{v_p}{|\mc S|}\,,
			\ee
for every action $i$ in $\mc S$ and $p$ in $\mc P$.
	\end{enumerate}
\end{corollary}
\begin{IEEEproof}
The statement follows from Proposition \ref{prop:graph} and Theorem \ref{thm:conv_global}.
\end{IEEEproof}
\begin{example}\label{ex:3}
	\begin{figure}
		\centering	
		\begin{tikzpicture}[scale = 1.5]
			\node[draw, circle] (1) at (0,0) {o};
			\node[draw, circle] (2) at (2,0) {a};
			\node[draw, circle] (3) at (4,0) {d};
			\path (1) edge [bend right=-30, ->]
			node [above] {$e_{1}$} (2);
			\path (1) edge [bend right=30, ->]
			node [below] {$e_{2}$} (2);	
			\path (2) edge [bend right=-30, ->]
			node [above] {$e_{3}$} (3);
			\path (2) edge [bend right=30, ->]
			node [below] {$e_{4}$} (3);
		\end{tikzpicture}\\[5pt]
		\includegraphics[width=8cm]{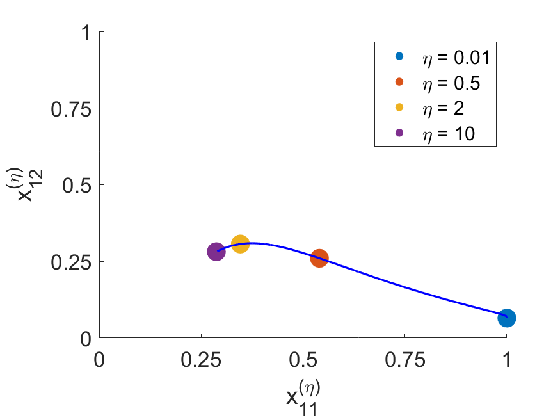}
		\caption{\label{fig:net}\emph{Top:} The multigraph of the routing game of Example \ref{ex:3}. \emph{Bottom:} Two components of the curve of globally asymptotically stable logit equilibria as a function of the noise level $\eta$.}
	\end{figure}
		Consider the routing game on the multigraph $\mc G = (\mc N, \mc E)$ in Figure \ref{fig:net} with link-route incidence matrix
		$$
		L = \left(\begin{matrix}
			1 & 1 & 0 & 0 \\
			0 & 0 & 1 & 1 \\
			1 & 0 & 1 & 0 \\
			0 & 1 & 0 & 1
		\end{matrix}\right)\,.
		$$ 
		Let $|\mc P| = 2$, $\theta = \ones$, and consider the link cost functions $\tau_{ep}(y_e) = p\cdot y_e + e$ for every $e$ in $\mc E$ and $p$ in $\mc P$.
		The game has a convex set of Nash equilibria
	$$
	\mc X^* = \left\{x^* \in \mc X: x^* = \left(\begin{matrix}
	1 & a \\
	0 & 1/4 - a \\
	0 & 1/4 - a \\
	0 & 1/2 + a
	\end{matrix}\right), \ 0 \le a \le \frac 14\right\}\,.
	$$
	Since the network is a series composition of two parallel multigraphs and all links are available to all populations, the logit dynamics are known to admit a globally asymptotically stable equilibrium $x^{(\eta)}$ due to Corollary \ref{cor:routing}.
	Figure \ref{fig:net} illustrates the entries $x_{11}^{(\eta)}$ and $x_{12}^{(\eta)}$ of such a curve. Notice that, in the vanishing noise limit, the logit equilibria approach one of the infinitely many Nash equilibria. While in Example \ref{ex:coordination} the set of limit logit equilibria $\mc X^{(0)}$ coincided with the set of Nash equilibria $\mc X^{*}$, in this example $\mc X^{(0)}$ \tcb{is a singleton and $\mc X^*$ contains infinitely many elements, hence $\mc X^{(0)}$} is a strict subset of the set of Nash equilibria $\mc X^*$. On the other hand, in the large noise limit, the logit equilibrium converges to the configuration where all populations are equally split on the available actions.
   \end{example}



\section{Conclusion}
\label{sec:evo_conclusion}
In this paper, we have studied the asymptotic stability of the logit dynamics in population games. First, we have proved the local asymptotic stability of strict Nash equilibria for low noise levels. Second, we have characterized the behavior of the logit dynamics for large noise levels, proving the existence of a globally asymptotically stable logit equilibrium. These results suggest the emergence of bifurcations for the logit dynamics in population games admitting multiple strict Nash equilibria, which have been also observed in the simulations. We then proved the global asymptotic stability of the logit dynamics in monotone separable population games, extending the class of games where the logit dynamics are known to admit globally asymptotically stable equilibria.

Several issues deserve future investigation. In particular, it would be worth rigorously analyzing the emergence of bifurcations for the logit dynamics as the noise level varies. 


%

\bibliographystyle{IEEEtran}
\bibliography{references.bib}

\begin{thebibliography}{10}
\providecommand{\url}[1]{#1}
\csname url@samestyle\endcsname
\providecommand{\newblock}{\relax}
\providecommand{\bibinfo}[2]{#2}
\providecommand{\BIBentrySTDinterwordspacing}{\spaceskip=0pt\relax}
\providecommand{\BIBentryALTinterwordstretchfactor}{4}
\providecommand{\BIBentryALTinterwordspacing}{\spaceskip=\fontdimen2\font plus
\BIBentryALTinterwordstretchfactor\fontdimen3\font minus
  \fontdimen4\font\relax}
\providecommand{\BIBforeignlanguage}[2]{{%
\expandafter\ifx\csname l@#1\endcsname\relax
\typeout{** WARNING: IEEEtran.bst: No hyphenation pattern has been}%
\typeout{** loaded for the language `#1'. Using the pattern for}%
\typeout{** the default language instead.}%
\else
\language=\csname l@#1\endcsname
\fi
#2}}
\providecommand{\BIBdecl}{\relax}
\BIBdecl

\bibitem{cianfanelli2019stability}
L.~Cianfanelli and G.~Como, ``On stability of users equilibria in heterogeneous
  routing games,'' in \emph{2019 IEEE 58th Conference on Decision and Control
  (CDC)}.\hskip 1em plus 0.5em minus 0.4em\relax IEEE, 2019, pp. 355--360.

\bibitem{cianfanelli2022stability}
L.~Cianfanelli, G.~Como, and T.~Toso, ``Stability and bifurcations in
  transportation networks with heterogeneous users,'' in \emph{2022 IEEE 61st
  Conference on Decision and Control (CDC)}.\hskip 1em plus 0.5em minus
  0.4em\relax IEEE, 2022, pp. 6371--6376.

\bibitem{blume2018population}
L.~E. Blume, ``Population games,'' in \emph{The economy as an evolving complex
  system II}.\hskip 1em plus 0.5em minus 0.4em\relax CRC Press, 2018, pp.
  425--460.

\bibitem{sandholm2010population}
W.~H. Sandholm, \emph{Population games and evolutionary dynamics}.\hskip 1em
  plus 0.5em minus 0.4em\relax MIT press, 2010.

\bibitem{smith1974theory}
J.~M. Smith, ``The theory of games and the evolution of animal conflicts,''
  \emph{Journal of theoretical biology}, vol.~47, no.~1, pp. 209--221, 1974.

\bibitem{schmeidler1973equilibrium}
D.~Schmeidler, ``Equilibrium points of nonatomic games,'' \emph{Journal of
  statistical Physics}, vol.~7, no.~4, pp. 295--300, 1973.

\bibitem{milchtaich2005topological}
I.~Milchtaich, ``Topological conditions for uniqueness of equilibrium in
  networks,'' \emph{Mathematics of Operations Research}, vol.~30, no.~1, pp.
  225--244, 2005.

\bibitem{roughgarden2002bad}
T.~Roughgarden and {\'E}.~Tardos, ``How bad is selfish routing?'' \emph{Journal
  of the ACM (JACM)}, vol.~49, no.~2, pp. 236--259, 2002.

\bibitem{roughgarden2003price}
T.~Roughgarden, ``The price of anarchy is independent of the network
  topology,'' \emph{Journal of Computer and System Sciences}, vol.~67, no.~2,
  pp. 341--364, 2003.

\bibitem{roughgarden2004bounding}
T.~Roughgarden and {\'E}.~Tardos, ``Bounding the inefficiency of equilibria in
  nonatomic congestion games,'' \emph{Games and economic behavior}, vol.~47,
  no.~2, pp. 389--403, 2004.

\bibitem{patriksson2015traffic}
M.~Patriksson, \emph{The traffic assignment problem: models and methods}.\hskip
  1em plus 0.5em minus 0.4em\relax Courier Dover Publications, 2015.

\bibitem{dafermos1973toll}
S.~C. Dafermos, ``Toll patterns for multiclass-user transportation networks,''
  \emph{Transportation science}, vol.~7, no.~3, pp. 211--223, 1973.

\bibitem{weibull1997evolutionary}
J.~W. Weibull, \emph{Evolutionary game theory}.\hskip 1em plus 0.5em minus
  0.4em\relax MIT press, 1997.

\bibitem{hofbauer2003evolutionary}
J.~Hofbauer and K.~Sigmund, ``Evolutionary game dynamics,'' \emph{Bulletin of
  the American mathematical society}, vol.~40, no.~4, pp. 479--519, 2003.

\bibitem{blume1993statistical}
L.~E. Blume, ``The statistical mechanics of strategic interaction,''
  \emph{Games and economic behavior}, vol.~5, no.~3, pp. 387--424, 1993.

\bibitem{marden2012revisiting}
J.~R. Marden and J.~S. Shamma, ``Revisiting log-linear learning: Asynchrony,
  completeness and payoff-based implementation,'' \emph{Games and Economic
  Behavior}, vol.~75, no.~2, pp. 788--808, 2012.

\bibitem{mckelvey1995quantal}
R.~D. McKelvey and T.~R. Palfrey, ``Quantal response equilibria for normal form
  games,'' \emph{Games and economic behavior}, vol.~10, no.~1, pp. 6--38, 1995.

\bibitem{dafermos1972traffic}
S.~C. Dafermos, ``The traffic assignment problem for multiclass-user
  transportation networks,'' \emph{Transportation science}, vol.~6, no.~1, pp.
  73--87, 1972.

\bibitem{farokhi2013heterogeneous}
F.~Farokhi, W.~Krichene, A.~M. Bayen, and K.~H. Johansson, ``A heterogeneous
  routing game,'' in \emph{2013 51st Annual Allerton Conference on
  Communication, Control, and Computing (Allerton)}.\hskip 1em plus 0.5em minus
  0.4em\relax IEEE, 2013, pp. 448--455.

\bibitem{mehr2019will}
N.~Mehr and R.~Horowitz, ``How will the presence of autonomous vehicles affect
  the equilibrium state of traffic networks?'' \emph{IEEE Transactions on
  Control of Network Systems}, vol.~7, no.~1, pp. 96--105, 2019.

\bibitem{arcak2020dissipativity}
M.~Arcak and N.~C. Martins, ``Dissipativity tools for convergence to {N}ash
  equilibria in population games,'' \emph{IEEE Transactions on Control of
  Network Systems}, vol.~8, no.~1, pp. 39--50, 2020.

\bibitem{hofbauer2001nash}
J.~Hofbauer, ``From {N}ash and {B}rown to {M}aynard {S}mith: equilibria,
  dynamics and {ESS},'' \emph{Selection}, vol.~1, no. 1-3, pp. 81--88, 2001.

\bibitem{hofbauer2009stable}
J.~Hofbauer and W.~H. Sandholm, ``Stable games and their dynamics,''
  \emph{Journal of Economic theory}, vol. 144, no.~4, pp. 1665--1693, 2009.

\bibitem{hofbauer2007evolution}
------, ``Evolution in games with randomly disturbed payoffs,'' \emph{Journal
  of economic theory}, vol. 132, no.~1, pp. 47--69, 2007.

\bibitem{hofbauer2002global}
------, ``On the global convergence of stochastic fictitious play,''
  \emph{Econometrica}, vol.~70, no.~6, pp. 2265--2294, 2002.

\bibitem{fox2013population}
M.~J. Fox and J.~S. Shamma, ``Population games, stable games, and passivity,''
  \emph{Games}, vol.~4, no.~4, pp. 561--583, 2013.

\bibitem{gao2020passivity}
B.~Gao and L.~Pavel, ``On passivity, reinforcement learning, and higher order
  learning in multiagent finite games,'' \emph{IEEE Transactions on Automatic
  Control}, vol.~66, no.~1, pp. 121--136, 2020.

\bibitem{martinez2023distributed}
J.~Martinez-Piazuelo, C.~Ocampo-Martinez, and N.~Quijano, ``Distributed {N}ash
  equilibrium seeking in strongly contractive aggregative population games,''
  \emph{IEEE Transactions on Automatic Control}, vol.~69, no.~7, pp.
  4427--4442, 2023.

\bibitem{park2019population}
S.~Park, N.~C. Martins, and J.~S. Shamma, ``From population games to payoff
  dynamics models: A passivity-based approach,'' in \emph{2019 IEEE 58th
  Conference on Decision and Control (CDC)}.\hskip 1em plus 0.5em minus
  0.4em\relax IEEE, 2019, pp. 6584--6601.

\bibitem{park2018passivity}
S.~Park, J.~S. Shamma, and N.~C. Martins, ``Passivity and evolutionary game
  dynamics,'' in \emph{2018 IEEE Conference on Decision and Control
  (CDC)}.\hskip 1em plus 0.5em minus 0.4em\relax IEEE, 2018, pp. 3553--3560.

\bibitem{park2019payoff}
S.~Park, N.~C. Martins, and J.~S. Shamma, ``Payoff dynamics model and
  evolutionary dynamics model: Feedback and convergence to equilibria,''
  \emph{arXiv preprint arXiv:1903.02018}, 2019.

\bibitem{FB-CTDS}
\BIBentryALTinterwordspacing
F.~Bullo, \emph{Contraction Theory for Dynamical Systems}, {1.1}~ed.\hskip 1em
  plus 0.5em minus 0.4em\relax Kindle Direct Publishing, 2023. [Online].
  Available: \url{https://fbullo.github.io/ctds}
\BIBentrySTDinterwordspacing

\bibitem{Brock.Durlauf:2001}
W.~Brock and S.~Durlauf, ``Discrete choice with social interactions,''
  \emph{Review and Economic Studies}, vol.~68, pp. 235--260, 2001.

\bibitem{Kurtz1981}
T.~G. Kurtz, \emph{Approximation of population processes}.\hskip 1em plus 0.5em
  minus 0.4em\relax Philadelphia: SIAM, 1981, vol.~36.

\bibitem{rudin1976principles}
W.~Rudin \emph{et~al.}, \emph{Principles of mathematical analysis}.\hskip 1em
  plus 0.5em minus 0.4em\relax McGraw-hill New York, 1976, vol.~3.

\bibitem{lovisari2014stability}
E.~Lovisari, G.~Como, and K.~Savla, ``Stability of monotone dynamical flow
  networks,'' in \emph{53rd IEEE Conference on Decision and Control}.\hskip 1em
  plus 0.5em minus 0.4em\relax IEEE, 2014, pp. 2384--2389.

\bibitem{Como.Lovisari.Savla:2015}
G.~Como, E.~Lovisari, and K.~Savla, ``Throughput optimality and overload
  behavior of dynamical flow networks under monotone distributed routing,''
  \emph{IEEE Transactions on Control of Network Systems}, vol.~2, no.~1, pp.
  57--67, 2015.

\bibitem{como2017resilient}
G.~Como, ``On resilient control of dynamical flow networks,'' \emph{Annual
  Reviews in Control}, vol.~43, pp. 80--90, 2017.

\bibitem{Margaliot:12}
M.~Margaliot and T.~Tuller, ``Stability analysis of the ribosome flow model,''
  \emph{IEEE/ACM Transactions on Computational Biology and Bioinformatics},
  vol.~9, no.~5, pp. 1545--1552, 2012.

\bibitem{beckmann1956studies}
M.~Beckmann, C.~B. McGuire, and C.~B. Winsten, ``Studies in the economics of
  transportation,'' Tech. Rep., 1956.

\bibitem{border1985fixed}
K.~C. Border, \emph{Fixed point theorems with applications to economics and
  game theory}.\hskip 1em plus 0.5em minus 0.4em\relax Cambridge University
  Press, 1985.

\bibitem{jafarpour2021weak}
S.~Jafarpour, P.~Cisneros-Velarde, and F.~Bullo, ``Weak and semi-contraction
  for network systems and diffusively coupled oscillators,'' \emph{IEEE
  Transactions on Automatic Control}, vol.~67, no.~3, pp. 1285--1300, 2021.

\end{thebibliography}

\appendices

\section{Proof of Proposition \ref{prop:existenceNash}}\label{app:prop1}
(i) See \cite[Theorem 2.1.1]{sandholm2010population} for the existence of at least an equilibrium. Since $\mc X^* \subseteq \mc X$ and $\mc X$ is bounded, $\mc X^*$ is bounded too. Moreover, since the reward function $r$ is continuous and $\mc X$ is closed, \eqref{eq:ess} implies that $\mc X^*$ is close. Closeness and boundedness imply compactness of $\mc X^*$.

(ii) Consider a non-monomorphic equilibrium $\wz$. Hence, there exists a population $p$ in $\mc P$ and two actions $i,j$ in $\mc S_p$ such that $x^*_{ip},x^*_{jp}>0$. By definition of Nash equilibrium, this implies that $r_{ip}(x^*) \ge r_{jp}(x^*)$ and $r_{jp}(x^*) \ge r_{ip}(x^*)$, hence $r_{ip}(x^*)=r_{jp}(x^*)$, proving that $x^*$ is not strict. Hence, all strict equilibria are monomorphic.

(iii) Consider a strict Nash equilibrium $\wz_{ip} = v_p \delta^{(s_p)}_i$. By definition of strict equilibrium and by continuity of the reward function, there exists a small enough $\epsilon>0$ such that for all $p$ in $\mc P$
$$
r_{s_p p}(x) > r_{ip}(x), \quad \forall i \in \mc S_p \setminus \{s_p\}, \forall x \in \mc B_{\epsilon}(\wz)\,.
$$
Consider now a Nash equilibrium $\tilde x$ in $\mc B_\epsilon(\wz)$. Hence, by continuity of the reward function, $r_{s_p p}(\tilde x) > r_{ip}(\tilde x)$ for every $i$ in $\mc S_p \setminus \{s_p\}$. Since $\tilde x$ is a Nash equilibrium, this implies that $\tilde x_{ip}=0$ for every $i$ in $\mc S_p \setminus \{s_p\}$, proving that necessarily $\tilde x = \wz$. $\qed$

\section{Proof of Lemma \ref{lemma0}}\label{app:0}
\tcb{(i) The right-hand side of equation \eqref{def:G} is nonnegative for every action $i$ in $\mc S$. Moreover, $$\sum_{i\in\mc S}F_{ip}(x,\eta)=v_p\,,$$ for every configuration $x$ in $\mc X$ and population $p$ in $\mc P$. Furthermore, the presence of the factor $\1_{\mc S_p}(i)$ in the right-hand side of \eqref{def:G} ensures that $F_{ip}(x,\eta)=0$ for every action $i$ in $\mc S\setminus\mc S_p$. Hence, $F(x,\eta)\in\mc X$ for every configuration $x$ in $\mc X$.}

\tcb{(ii) The statement follows from the fact that
$$
F_{ip}(x) = \frac{v_p}{1+\sum\limits_{j \in \mc S_p \setminus \{i\}} \exp((r_{jp}(x)-r_{ip}(x))/\eta)} \stackrel{\eta \to +\infty}{\longrightarrow} \frac{v_p}{|\mc S_p|} 
$$
for every $x$ in $\mc X$, $p$ in $\mc P$, and $i$ in $\mc S_p$.}

\tcb{(iii) If an action $i$ does not belong to $\mc S^*_p(x)$, then $r_{ip}(x) < r_{jp}(x)$ for some $j$ in $\mc S_p \setminus \{i\}$. Hence,
$$
	\ds F_{ip}(x) = \frac{v_p}{1+\sum\limits_{j \in \mc S_p \setminus \{i\}} \exp((r_{jp}(x)-r_{ip}(x))/\eta)} \stackrel{\eta \to 0^+}{\longrightarrow} 0\,. 
$$ 
	On the other hand, if $i$ belongs to $\mc S^*_p(x)$, then
	$$
	\ds F_{ip}(x) = \frac{v_p}{1+ \!\! \sum\limits_{j \in \mc S_p \setminus \{i\}} \exp((r_{jp}(x)-r_{ip}(x))/\eta)} \stackrel{\eta \to 0^+}{\longrightarrow} \frac{v_p}{|\mc S^*_p(x)|}\,.
	$$
The two relations together prove the statement.
}

\tcb{(iv) For every noise level $\eta>0$, population $p$ in $\mc P$, and actions $i$ and $j$ in $\mc S_p$, the function $h_{ij}(x)=\exp((r_{jp}(x)-r_{ip}(x))/\eta)$ is Lipschitz on the configuration space $\mc X$, as it is the composition of the exponential function (which is locally Lipschitz) with the difference of the reward function $r_{jp}(x)-r_{ip}(x)$, which are assumed to be Lipschitz on the compact configuration space $\mc X$. 
It follows that the function 
$$x\mapsto F_{ip}(x,\eta)=\frac{v_p}{\ds 1+\sum_{j\in\mc S_p\setminus\{i\}}h_{ij}(x)}$$ 
is Lipschitz on $\mc X$, as it is the composition of the function $y\mapsto 1/(1+y)$, which is Lipschitz on $\R_+$, with 
the sum $\sum_{j\in\mc S_p\setminus\{i\}}h_{ij}(x)$ of Lipschitz functions. Hence, for every noise level $\eta>0$, the function $x\mapsto F(x,\eta)$ defined by \eqref{def:G} is Lipschitz on the configuration space $\mc X$. By the same arguments, the fact that the reward function is assumed continuously differentiable on the interior of $\mc X$, together with the continuous differentiability of the exponential function on $\R$ and of $y \mapsto 1/(1+y)$ on $\R_+$, implies that the function  $x\mapsto F(x,\eta)$ is continuously differentiable on the interior of $\mc X$.}

\tcb{(v) The proof relies on the fact that the exponential function is continuously differentiable and follows the same steps as in the proof of point (iv).}

\tcb{(vi) For every noise level $\eta>0$, point (i) implies that $x\mapsto F(x,\eta)$ maps the non-empty compact convex set $\mc X$ to itself. Point (iv) implies that $x \mapsto F(x,\eta)$ is continuous. Hence, Brouwer's fixed point theorem \cite{border1985fixed} guarantees the existence of at least one fixed point $x^*=F(x^*,\eta)$ in $\mc{X}$.} $\qed$

\section{Proof of Lemma \ref{lemma:existence-uniqueness}}\label{sec:proof-lemma:existence-uniqueness}
\tcb{(i) Lemma \ref{lemma0}(iv) ensures that the function $x \mapsto F(x,\eta)$ is Lipschitz. This ensures the local existence and uniqueness of a solution $x(t)$ of the logit dynamics \eqref{eq:logit} for every initial configuration $x(0)$ in $\mc X$.
Moreover, Lemma \ref{lemma0}(i) implies that, for every $\eta>0$, $F(x,\eta)$ belongs to $\mc X$ for every $x$ in $\mc X$, hence in particular $F_{ip}(x) = 0$ for every $p$ in $\mc P$ and $i$ in $\mc S \setminus \mc S_p$. This in turn implies that for every $x$ in $\mc X$,
\be\label{eq:invariant} 
\dot{x}_{ip} = F_{ip}(x) - x_{ip} = 0\,, \quad \forall p \in \mc P\,, \ \forall i \in \mc S \setminus \mc S_p\,,\ee
and
\be\label{eq:invariant2}
\sum_{i \in \mc S} \dot x_{ip} = \sum_{i \in \mc S} (F_{ip}(x) - x_{ip}) = 0\,, \quad \forall p \in \mc P\,.
\ee
Furthermore, since $F(x)$ is non-negative for every $x$ in $\mc X$, it follows from \eqref{eq:logit} that, if $x_{ip}=0$, then $\dot x_{ip} \ge 0$.}
This, together with \eqref{eq:invariant}-\eqref{eq:invariant2}, implies that the configuration space $\mc X$ is invariant under the logit dynamics \eqref{eq:logit}. Since the configuration space $\mc X$ is compact, existence and uniqueness of a solution $x(t)$ of \eqref{eq:logit} are ensured globally.

(ii) Notice that logit equilibria coincide with fixed points of $x \mapsto F(x,\eta)$. Hence, Lemma \ref{lemma0}(vi) implies the set of logit equilibria $\mc X^{(\eta)}$ is non-empty for every noise level $\eta>0$. Notice that $\mc X^{(\eta)}$ is a bounded set because it is a subset of the configuration space $\mc X$, which is a compact subset of $\R^{\mc S\times\mc P}$. Moreover, continuity of the function $x\mapsto F(x,\eta)$ for every $\eta>0$ \tcb{(c.f. Lemma \ref{lemma0}(iv))} implies that the set of its fixed points $\mc X^{(\eta)}$ is closed. Therefore, the set of logit equilibria $\mc X^{(\eta)}$ is compact for every noise level $\eta>0$.

(iii) Since the configuration space $\mc X$ is a compact subset of $\R^{\mc S\times\mc P}$, every sequence of points in $\mc X$ admits a converging subsequence, so that the set $\mc X^{(0)}$ of limit logit equilibria is nonempty. Now, consider a limit logit equilibrium $\wz$ in $\mc X^{(0)}$, and let $\eta_n>0$ and $x^{n}$ in $\mc X^{(\eta_n)}$ be two sequences such that  $$\eta_n \stackrel{n \to +\infty}{\longrightarrow} 0\,,\qquad x^{n}\stackrel{n \to +\infty}{\longrightarrow} \wz\,.$$
Consider an action $i$ not in $\mc S^*_p(x^*)$, i.e., such that $$r_{jp}(\wz)>r_{ip}(\wz)\,,$$ for some $j$ in $\mc{S}_p$. Hence,
\begin{equation}
	\lim_{n \rightarrow +\infty} F_{ip}(x^n,\eta_n) = 0.
	\label{eq:lim}
\end{equation}
From \eqref{eq:logit} and \eqref{eq:lim} it follows
\begin{equation*}
	\wz_{ip}=\lim_{n \to +\infty}x^{n}_{ip}
	= \lim_{n \to +\infty} F_{ip}(x^n,\eta_n)=  
	0.
\end{equation*}
Hence, every limit logit equilibrium $\wz$ is a Nash equilibrium, i.e., $\mc X^{(0)}\subseteq\mc X^*$.
To prove that $\mc X^{(0)}$ is closed, consider a sequence $x^k$ in $\mc X^{(0)}$ such that $x^k \stackrel{k \to +\infty}{\longrightarrow} \wz$ in $\mc X$. 
By definition of $\mc X^{(0)}$, for every $k$, there exist two converging sequences $\eta_m>0$  and $x^{(k,m)}$  in $\mc X^{(\eta_m)}$ such that $\eta_m \stackrel{m \to +\infty}{\longrightarrow} 0$ and $x^{(k,m)} \stackrel{m \to +\infty}{\longrightarrow} x^k$.
This implies the existence of a sequence of positive integers $m_k$ such that
$$
|x^{(k,m_k)}-x^k| < \frac{1}{k}\,,\qquad \forall k\ge1\,.
$$
We now prove that $x^{(k,m_k)} \stackrel{k \to +\infty}{\longrightarrow} \wz$. For every $\epsilon>0$, we take $k$ such that $|x^k - \wz| < \epsilon/2$ and $k>2/\epsilon$. Then,
\begin{equation}
	\label{eq:omega0}
	|x^{(k,m_k)} - \wz| \le |x^{(k,m_k)} - x^k| + |x^k - \wz| < \frac{1}{k} + \frac{\epsilon}{2} < \epsilon.
\end{equation}
Since $x^{(k,m_k)}$ is by construction an element of $\mc X^{(\eta_{m_k})}$, \eqref{eq:omega0} shows that $\wz$ is an accumulation point of logit equilibria and thus is contained in $\mc X^{(0)}$. This proves that $\mc X^{(0)}$ is closed. Since $\mc X^{(0)}\subseteq\mc X$ is bounded, this implies that  $\mc X^{(0)}$ is compact, thus completing the proof. $\qed$

\section{Proof of Lemma \ref{lemma:G}} \label{app:G}
	\emph{(i)}
	Let $x^*$ in $\mc X_s^*$ be a strict Nash equilibrium and, for every $p$ in $\mc{P}$, let $s_p$ denote the action in use by population $p$, so that $\wz_{ip} = v_p \delta_i^{(s_p)}$. Let
$$\alpha := \min\left\{r_{s_pp}(\wz)-r_{j p}(\wz):\,p \in \mc{P},j \in \mc{S}_p \setminus \{s_p\}\right\}\,.
$$
	Since $\wz$ is a strict Nash equilibrium, we have that $\alpha>0$. Let
	\begin{equation*}
		\bar{\epsilon}=\max\Big\{\epsilon\! >\! 0: \!\min_{x \in \mc O_\epsilon(\wz)} \min_{p \in \mc{P}}\min_{j \ne s_p} (r_{s_pp}(x)-r_{j p}(x)) \ge  \frac{\alpha}{2}\!\Big\}\,.
	\end{equation*} 
	
	Observe that $\bar{\epsilon}>0$, since $x^*$ is a strict Nash equilibrium and the reward function is continuous. Observe also that for every $\epsilon$ in $(0, \bar{\epsilon}]$ and population $p$ in $\mc P$, the action $s_p$ is strictly optimal for every $x$ in $\mc O_\epsilon(\wz)$ by definition of $\bar{\epsilon}$, i.e.,
	\begin{equation}
		\label{eq:strict_intorno}
		r_{s_p p}(x) > r_{jp}(x) \quad \forall x \in \mc O_\epsilon(\wz), p \in \mc{P}, j \in \mc{S}_p \setminus \{s_p\}. 
	\end{equation}
	Thus, for every $\epsilon$ in $(0, \bar{\epsilon}]$, $x$ in $\mc O_\epsilon(\wz)$ and $p$ in $\mc{P}$,
	\begin{equation}
		\label{eq:lim_eta_}
		\lim_{\eta \to 0^+}F_{ip}(x,\eta) = 
		v_p \delta^{(s_p)}_i = \wz_{ip}\,.
	\end{equation}
	Since $$\mc O_\epsilon(\wz) \xrightarrow{\eps \to 0^+} \{x^*\}\,,$$ this implies by continuity of $\eta \mapsto F(x,\eta)$ (\tcb{c.f. Lemma \ref{lemma0}(v)}) that, for every $\epsilon$ in $(0,\bar{\epsilon}]$, there exists a small enough $\eta_\epsilon>0$ such that for $\eta$ in $(0,\eta_\epsilon]$, $x \mapsto F(x,\eta)$ maps $\mc O_\epsilon(\wz)$ to itself.
	
	\emph{(ii)} The continuity of $\bar F$ follows from \tcb{the continuity of $F$ (c.f. Lemma \ref{lemma0}(iv)-(v))} and from \eqref{eq:lim_eta_}, which implies $$\lim_{\eta \to 0^+} \bar F(x,\eta) = \wz\,,\qquad\forall x\in\mc O_{\bar\epsilon}(\wz)\,.$$ 
	Since $F$ is continuously differentiable (\tcb{c.f. Lemma \ref{lemma0}(iv)-(v)}), we need to investigate the differentiability of $\bar F$ in $\eta=0$ only. We first show that $\bar{F}(x,\eta)$ is continuously differentiable with respect to $x$, which is equivalent to proving \eqref{eq:J}.
	Let $$\Delta_{ij}^p(x) = r_{ip}(x)-r_{jp}(x)\,.$$
	Hence,
	$$
\frac{\partial F_{ip} (x,\eta)}{\partial x_{jq}}=
\frac{\ds v_pe^{r_{ip}(x)/\eta}\sum\nolimits_{k\ne i}
	\frac{\ds\partial \Delta_{ik}^p(x)}{\ds\partial x_{jq}} e^{r_{kp}(x)/\eta}}{\ds\eta\left(\sum\nolimits_{k} e^{r_{kp}(x)/\eta}\right)^2}.
	$$
	Now, we split the analysis into two parts, depending on whether $i = s_p$ or not. For every $p,q$ in $\mc{P}$, $i$ in $\mc{S}_p \setminus \{s_p\}$, $k$ in $\mc{S}_p \setminus \{i\}$ and $x$ in $\mc O_{\bar\epsilon}(\wz)$, it follows from \eqref{eq:strict_intorno} that
	\begin{equation}\label{eq:jac4}
		\lim_{\eta \to 0^+} \frac{\ds e^{r_{ip}(x)/\eta}e^{r_{lp}(x)/\eta}}{\ds \eta(\sum\nolimits_{k} e^{r_{kp}(x)/\eta})^2} = 0\,,
	\end{equation}
	implying for every $x$ in $\mc O_{\bar\epsilon}(\wz)$
	\begin{equation}
		\label{eq:piece1}
		\lim_{\eta \to 0^+}\frac{\partial F_{ip} (x,\eta)}{\partial x_{jq}} = 0, \ \forall p,q \in \mc{P}, i \in \mc{S}_p \setminus \{s_p\}, j \in \mc{S}_q.
	\end{equation}
	If instead $i = s_p$ and $k$ is in $\mc S_p \setminus i$, \eqref{eq:jac4} with $i=s_p$, we have that for every $p,q$ in $\mc{P}$ and $x$ in $\mc O_{\bar\epsilon}(\wz)$,
	$$
	\lim_{\eta \to 0^+} \frac{\ds e^{r_{s_p p}(x)/\eta}e^{r_{kp}(x)/\eta}}{\eta\left(\ds \sum\nolimits_{k \in \mc S_p} e^{r_{kp}(x)/\eta}\right)^2} = 0,
	$$
	hence
	\begin{equation}
		\label{eq:piece2}
		\! \lim_{\eta \to 0^+} \frac{\partial F_{s_p p}  (x,\eta)}{\partial x_{jq}} = 0\,, \forall p,q \in \mc{P},\ j \in \mc{S}_q, x \in \mc O_{\bar\epsilon}(\wz).
	\end{equation}
	Eq. \eqref{eq:piece1} and \eqref{eq:piece2} together prove \eqref{eq:J} and the fact that $\bar{F}(x,\eta)$ is continuously differentiable with respect to $x$ for every $x$ in $\mc O_{\bar \epsilon}(\wz)$ and $\eta \ge 0$. 
	We now prove the continuous differentiability of $\bar F(x,\eta)$ with respect to $\eta$ for every $x$ in $\mc O_{\bar{\epsilon}}(\wz)$ and $\eta \ge 0$. 
	In particular, we aim to prove that
	$$
	\lim_{\eta \to 0^+}\frac{\partial F_{ip}(x,\eta)}{\partial \eta} = 0, \quad \forall p \in \mc P, i \in \mc S_p, x \in \mc O_{\bar\epsilon}(\wz).
	$$ 
	Direct computation yields
	\begin{equation}
		\label{eq:der_eta}
		\frac{\partial F_{ip}(x,\eta)}{\partial \eta} = \frac{\ds v_p \sum\nolimits_{k \neq i} \Delta_{ki}^p(x) e^{\Delta_{ki}^p(x)/\eta}}{\ds \eta^2(1+\sum\nolimits_{k \neq i} e^{\Delta_{ki}^p(x)/\eta})^2}.
	\end{equation}
	We split the analysis into two parts. On the one hand, if $i = s_p$, observe that $\Delta_{ks_p}^p(x)>0$ for every $k$ in $\mc S_p \setminus \{s_p\}$ and $x$ in $\mc O_{\bar \epsilon}(\wz)$. Hence, by \eqref{eq:der_eta},
$$
		\lim_{\eta \to 0^+}\frac{\partial F_{s_p p}(x,\eta)}{\partial \eta} = 0, \quad \forall p \in \mc{P}, x \in \mc O_{\bar\epsilon}(\wz).
$$
	On the other hand, for every $i$ in $\mc S_p \setminus \{s_p\}$ we get that, as $\eta \to 0^+$, the numerator and the denominator in \eqref{eq:der_eta} are dominated by the term with $k = s_p$, and $\Delta^p_{s_pi}(x) > 0$ for every $x$ in $\mc O_{\bar\epsilon}(\wz)$. Hence,
$$
		\lim_{\eta \to 0^+}\frac{\partial F_{ip}(x,\eta)}{\partial \eta} = 0 \quad \forall p \in \mc P, \forall i \in \mc{S}_p \setminus \{s_p\}, x \in \mc O_{\bar\epsilon}(\wz).
$$
	This shows that $\eta \mapsto \bar{F}(x,\eta)$ is continuously differentiable on $(0,+\infty)$ and concludes the proof. $\qed$
	
	\section{Proof of Lemma \ref{prp:contractivity}}\label{sec:proof-prp-contractivity}
	For simplicity of notation, we omit the dependence on $t$.	
	Let $h=x-\tilde x$. 
	By definition of the \tcb{$l_1$-norm} and linearity of the derivative, we get 
	\begin{equation*} 
		\ba{rcl}
			\ds\frac{\de}{\de t}\ds\tcb{\|h\|}   
			&= &\ds\frac{\de}{\de t} \sum_{i=1}^n |h_i|\\
			&= &\ds\sum_{i=1}^n\sigma_i(\dot{x_i}-\dot{\tx}_i)\\
			&= &\ds\sum_{i=1}^n\sigma_i(f_i(x)-f_i(\tx))\\
			&= &\ds\sum_{i=1}^n\sigma_i(f_i(\tx+h)-f_i(\tx))\,,
		\ea
	\end{equation*}  
	where $\sigma_i=\sign (h_i)$. 
	Observe that  
\be\label{observe}
\ba{rcl}
\ds f_i(\tx+h)-f_i(\tx)
&=&\ds \int_0 ^1 \frac{\de f_i(\tx+\tau h)}{\de\tau} \de\tau\\ 
&=&\ds\int_0 ^1 \sum_{j=1}^n{\frac{\partial{f_i}}{\partial{x_j}}}(\tx+\tau h)h_j \de\tau\\
&=&\ds \sum_{j=1}^nh_j \int_0 ^1g_i^j(\tau)\de\tau\,,\ea
\ee
where $\ds g_i^j(\tau)={\frac{\partial{f_i}}{\partial{x_j}}}(\tx+\tau h)$. 
Therefore, we get that 
$$
		\ba{rcl}
		\ds\frac{\de}{\de t}\tcb{\|h\|}  
		&=&\ds \sum_{i=1}^n\sigma_i \sum_{j=1}^nh_j\int_0 ^1  g_i^j(\tau) \de\tau\\[10pt]
		&=&\ds \sum_{i=1}^n\left( \sum_{j \neq i}\sigma_ih_j \int_0 ^1g_i^j(\tau) \de\tau  + |h_i|\int_0 ^1  g_i^i(\tau) \de\tau\right)\\[10pt]
		&\le&\ds \sum_{i=1}^n\left( \sum_{j \neq i}|h_j| \int_0 ^1g_i^j(\tau) \de\tau  + |h_i|\int_0 ^1  g_i^i(\tau) \de\tau\right)\\[10pt]
         	&=&\ds \sum_{i=1}^n\left( \sum_{j=1}^n|h_j| \int_0 ^1g_i^j(\tau) \de\tau\right)\\[10pt]
	        	&=&\ds \sum_{j=1}^n|h_j|\int_0 ^1\sum_{i=1}^n{\frac{\partial{f_i}}{\partial{x_j}}}(\tx+\tau h) \de\tau\\[10pt]
		&\le&\ds  -\tcb{||h||} c \,.
		\ea
$$	It then follows from Gronwall's lemma that 
$$	\tcb{||x(t) - \tx(t)|| =||h(t)||\le e^{-ct}||h(0)||= e^{-ct}||x_0-\tx_0||}\,,$$
thus proving point (i) of the claim. We refer to \cite[Theorem 13]{jafarpour2021weak} for the proof of point (ii).
$\qed$

\section{Proof of Proposition \ref{prop:graph}}
\label{app:graph}
Let $\mc A_{k}$ denote the set of routes from $o^{(k)}$ to $d^{(k)}$, and notice that the route set of $\mc S$ is the cartesian product $\mc S = \prod_{k=1}^{l} \mc A_{k}$. Let $L^{(k)}$ in $\{0,1\}^{\mc E^{(k)} \times \mc A_{k}}$ denote the link-route incidence matrix of graph $\mc G^{(k)}$. Observe that links that do not belong to any route do not affect the reward functions of the game, hence we assume without loss of generality that every link $e$ in $\mc E^{(k)}$ belongs to a route in $\mc A_{k}$, i.e., $L^{(k)}\ones = \ones$ for every $k = 1,\cdots,l$, since $\mc G^{(k)}$ is a parallel graph.
For a link $e$ in $\mc E^{(k)}$, there exists a unique route in $\mc A_k$, denoted $a^{(k)}(e)$, such that 
	$L^{(k)}_{e a^{(k)}(e)} = 1$. Notice also that $L_{ei} = L^{(k)}_{ei_k}$. Hence, for a link $e$ in $\mc E^{(k)}$, it follows from \eqref{eq:reward_routing} and \eqref{eq:z} that
	\be\label{eq:y_series}
	\ba{rcl}
		y_e 
		&=&\ds \sum_{p \in \mc P}\sum_{i \in \mc S}L_{ei} x_{ip} \theta_p\\[10pt]
		& =&\ds \sum_{p \in \mc P} \sum_{i \in \mc S} L^{(k)}_{ei_k} x_{ip} \theta_p  \\[10pt]
		&=&\ds \sum_{p \in \mc P}\theta_p \sum_{i \in \mc S: i_k =  a^{(k)}(e)} x_{ip} \\[10pt]
		&=&\ds z^{(k)}_{a^{(k)}(e)}(x)\,.\ea
	\ee
	Now, for every population $p$ in $\mc P$, $k = 1,\cdots,l$ and $a$ in $\mc A_{k}$ consider the function $$\varphi_{ap}^{(k)}(z_a^{(k)}) = - \sum_{e \in \mc E^{(k)}} L_{ea}^{(k)} \tau_{ep}(z_a^{(k)})\,,$$ which is non-increasing since the cost functions $\tau_{ep}$ are non-decreasing. Hence, from \eqref{eq:reward_routing} we get that 
	$$\ba{rcl}
		r_{ip}(x) &=&
		\ds - \sum_{e \in \mc E}L_{ei} \tau_{ep} (y_e(x)) \\[10pt]
		&=&\ds - \sum_{k = 1}^{l} \sum_{e \in \mc E^{(k)}} L^{(k)}_{ei_k} \tau_{ep}(y_e(x))\\[10pt]
		& =&\ds - \sum_{k = 1}^{l} \sum_{e \in \mc E^{(k)}} L^{(k)}_{ei_k} \tau_{ep}(z^{(k)}_{i_k}(x)) \\[10pt]
		&=&\ds \sum_{k = 1}^l \varphi_{i_k p}^{(k)}(z_{i_k}^{(k)})\,,\ea
$$
	where the second equality follows from the fact that the link sets $\mc E^{(k)}$ are disjoint (c.f. Definition \ref{def:series}), and the third one from \eqref{eq:y_series} and from the fact that $L_{ei_k}^{(k)} \neq 0$ implies $i_k =a^{(k)}(e)$, so that $$ L^{(k)}_{ei_k} \tau_{ep}(y_e(x)) =  L^{(k)}_{ei_k} \tau_{ep}(z^{(k)}_{a^{(k)}(e)}(x)) = L^{(k)}_{ei_k} \tau_{ep}(z^{(k)}_{i_k}(x))\,.$$ 
	We have thus proved that \eqref{eq:cost} holds true, so that the game is monotone separable. 
	$\qed$

\begin{IEEEbiography}[{\includegraphics[width=1in,height=1.25in,clip,keepaspectratio]{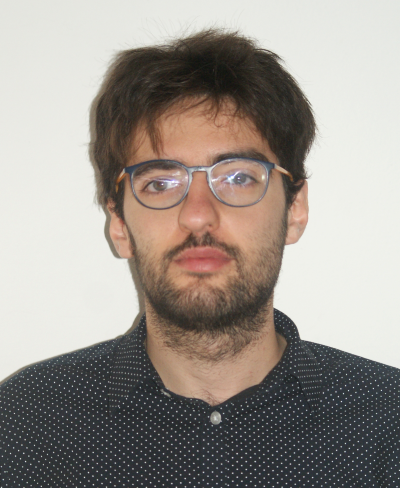}}]{Leonardo Cianfanelli} (M'21)
	received the B.Sc. (cum Laude) in physics and astrophysics in 2014 from Università degli Studi di Firenze, Italy, the M.S. in physics of complex systems (cum Laude) in 2017 from Università di Torino, Italy, and the PhD in pure and applied mathematics in 2022 from Politecnico di Torino. He is currently a research assistant at the Department  of  Mathematical Sciences, Politecnico di Torino, Italy. He was a Visiting Student at Laboratory for Information and Decision Systems, Massachusetts Institute of Technology, Cambridge, MA, USA, in 2018-2020. His research focuses on dynamics and control in network systems, with applications to transportation networks and epidemics.
\end{IEEEbiography}

\begin{IEEEbiography}[{\includegraphics[width=1in,height=1.25in,clip,keepaspectratio]{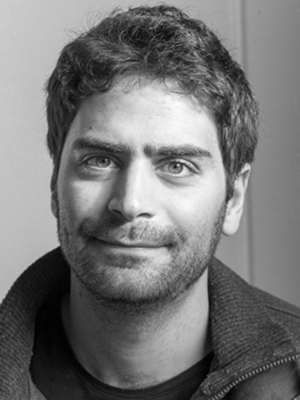}}]{Giacomo Como}(M'12) is  a  Professor at  the Department  of  Mathematical  Sciences,  Politecnico di  Torino,  Italy. He is also a Senior Lecturer  at  the  Automatic  Control  Department, Lund  University,  Sweden.  He  received the B.Sc., M.S., and Ph.D.~degrees in Applied Mathematics  from  Politecnico  di  Torino,  in  2002,  2004, and 2008, respectively. He was a Visiting Assistant in  Research  at  Yale  University  in  2006--2007  and  a Postdoctoral  Associate  at  the  Laboratory  for  Information  and  Decision  Systems,  Massachusetts  Institute of Technology in  2008--2011. Prof.~Como currently serves as Senior Editor for the \textit{IEEE Transactions on Control of Network Systems} and as Associate  Editor  for \textit{Automatica}. He served as Associate Editor for the  \textit{IEEE Transactions on Network Science and Engineering} (2015-2021) and for the \textit{IEEE Transactions on Control of Network Systems} (2016-2022).  
	He was  the  IPC  chair  of  the  IFAC  Workshop  NecSys'15,  a  semiplenary speaker  at  the  International  Symposium  MTNS'16, and the  chair  of the  {IEEE-CSS  Technical  Committee  on  Networks  and  Communications} in 2019--2024.  
	He  is  recipient  of  the 2015  George S.~Axelby  Outstanding Paper Award.  His  research interests  are in  dynamics,  information,  and  control  in  network  systems,  with  applications to  infrastructure,  social,  and  economic networks.
\end{IEEEbiography}


\end{document}